%% file: revised_main_draft.tex
 \documentclass[journal]{IEEEtran}

\usepackage{amsthm,amssymb,graphicx,multirow,amsmath,color,amsfonts}
\usepackage[update,prepend]{epstopdf}
\usepackage[noadjust]{cite}
\usepackage[latin1]{inputenc}
\usepackage{tikz}
\usetikzlibrary{arrows,calc}		
\usepackage{bbm} 
\usepackage{pdfpages}
\usepackage{tabulary}
\usepackage{multirow}
\usepackage{comment}
\usepackage{subcaption}
\usepackage[linesnumbered,ruled,vlined]{algorithm2e}
\pdfoutput=1
\usepackage{svg}
\usepackage{enumitem,kantlipsum}
\usepackage{balance}
\usepackage{tabularx}
\usepackage{makecell}               
\usepackage{array}                  

\def\chb#1{{\color{blue} #1}}
\renewcommand{\chb}[1]{\textcolor{black}{#1}}

\usepackage{xcolor}
\newenvironment{chbgrp}{\color{black}}{}

\everymath{\small} 

\allowdisplaybreaks 
\include{notation}

\begin{document}
\graphicspath{{./figures_pdf/}}
\title{
{\fontsize{20.8}{19}\selectfont Two-Stage Weighted Projection for Reliable Low-Complexity Cooperative and Non-Cooperative Localization}
}
\author{
Harish K. Dureppagari, R. Michael Buehrer, Harpreet S. Dhillon
\thanks{H. K. Dureppagari, R. M. Buehrer, and H. S. Dhillon are with Wireless@VT, Department of ECE, Virginia Tech, Blacksburg, VA 24061, USA. Email: \{harishkumard, rbuehrer, hdhillon\}@vt.edu. The support of the US NSF (Grants CNS-1923807 and CNS-2107276) and the US NIST (Grant CAN-70NANB22H070) is gratefully acknowledged. This paper has been published in part at the IEEE ICC Workshop 2025, Montreal, Canada~\cite{dureppagari2025icc}.
}
\vspace{-1mm}
}

\maketitle

\begin{abstract}
In this paper, we propose a two-stage weighted projection method (TS-WPM) for time-difference-of-arrival (TDOA)-based localization, providing provable improvements in positioning accuracy, particularly under high geometric dilution of precision (GDOP) and low signal-to-noise ratio (SNR) conditions. TS-WPM employs a two-stage iterative refinement approach that dynamically updates both range and position estimates, effectively mitigating residual errors while maintaining computational efficiency. Additionally, we extend TS-WPM to support cooperative localization by leveraging two-way time-of-arrival (TW-TOA) measurements, which enhances positioning accuracy in scenarios with limited anchor availability. To analyze TS-WPM, we derive its error covariance matrix and mean squared error (MSE), establishing conditions for its optimality and robustness. To facilitate rigorous evaluation, we develop a 3rd Generation Partnership Project (3GPP)-compliant analytical framework, incorporating 5G New Radio (NR) physical layer aspects as well as large-scale and small-scale fading. As part of this, we derive a generalized Cram{\'e}r-Rao lower bound (CRLB) for multipath propagation and introduce a novel non-line-of-sight (NLOS) bias model that accounts for propagation conditions and SNR variations. Our evaluations demonstrate that TS-WPM achieves near-CRLB performance and consistently outperforms state-of-the-art weighted nonlinear least squares (WNLS) in high GDOP and low SNR scenarios. Moreover, cooperative localization with TS-WPM significantly enhances accuracy, especially when an insufficient number of anchors (such as 2) are visible. Finally, we analyze the computational complexity of TS-WPM, showing its balanced trade-off between accuracy and efficiency, making it a scalable solution for real-time localization in next-generation networks.
\end{abstract}

\section{Introduction}
Localization is an essential component for numerous applications spanning navigation, autonomous systems, the Internet of Things (IoT), and next-generation communication networks like 5G Advanced and 6G. Over the years, significant advancements have been made in positioning methodologies, leveraging techniques such as TOA and TDOA-based localization. Among these, sophisticated methods such as nonlinear least squares (NLS)~\cite{kay1993fundamentals} and weighted NLS (WNLS)~\cite{buehrer2019handbook}, along with optimization techniques like semi-definite programming (SDP)~\cite{9376594,6268731,8241388} and multidimensional scaling (MDS)~\cite{10.1145/1138127.1138129,5342507} and machine learning-based techniques~\cite{Bhatti2018MachineLB,tian2023highprecisionmachinelearningbasedindoor} are recognized for their increased accuracy. However, they tend to be computationally intensive. In contrast, parallel projection methods (PPMs)~\cite{5683693_toa_coop_mod_ippm,5677549_tao_coop_ippm} are noted for their computational efficiency by avoiding complex operations like matrix inversion but are inferior to methods like NLS~\cite{10632801_harish_dyspan}. Note that PPMs have traditionally been confined to TOA-based localization, while SDP, MDS, and NLS methods are applicable to both TOA and TDOA-based localization. TOA-based localization requires precise time synchronization between anchors and user equipment (UEs), as well as awareness of signal transmission time. Conversely, TDOA-based localization does not require precise synchronization between anchors and UEs, is more resilient to clock offsets, and is more robust to multipath propagation. As a result, TDOA-based localization is often preferred, making methods like NLS, WNLS, SDP, and MDS often desirable. However, these methods may not be computationally feasible, especially for scenarios such as edge location computing (e.g., augmented reality (AR)/virtual reality (VR), drone navigation, and emergency geolocation)~\cite{10329418,10.1007/s00521-021-06815-9}, UE-driven non-terrestrial networks (NTN) access~\cite{dureppagari_ntn_10355106,dureppagari2024leo}, and power- and bandwidth-limited devices~\cite{10634076}. Motivated by these challenges, this paper proposes a novel two-stage weighted projection method (TS-WPM) for cooperative localization that provides provable gains in positioning accuracy while maintaining computational efficiency. Our proposed approach is particularly effective in challenging scenarios, such as low SNR or high GDOP environments \chb{(see Appendix~\ref{app::key_definitions} for the detailed definition of SNR and GDOP and the corresponding metrics used for evaluation purposes in this article)}.

\subsection{Related Work}\label{sec:related_work}
\begin{table*}[htbp]
\centering
\caption{\chb{Comparison of Related Work with the Proposed TS-WPM Approach}}
\label{tab:related_work_summary}
\renewcommand{\arraystretch}{1.5}
\setlength{\tabcolsep}{4pt}
\scriptsize
\begin{tabular}{|p{1.8cm}|p{6.8cm}|p{1.5cm}|p{6.6cm}|}
\hline
\textbf{Reference} & \textbf{Key Contributions} & \textbf{Scope} & \textbf{Limitations / Gaps} \\
\hline
Zhuo et al.~\cite{ZHUO2024} & Low-complexity TOA localization using greedy and list-based grid search methods & TOA-only & Heuristic approach with significant gap from CRLB; no support for cooperative localization \\
\hline
Wu et al.~\cite{5464262} & LS-based closed-form localization with improved stability under noise & TOA-only & No support for TDOA; does not account for SNR or multipath; no cooperative localization; lacks CRLB-based comparison \\
\hline
Liu et al.~\cite{8947090_liu_weight_ippm} & Weighted PPM for TOA localization using SNR-based gradient descent optimization & TOA-only & Not applicable to TDOA; requires synchronization between anchors and UEs; limited positioning accuracy; no cooperative localization \\
\hline
Liu et al.~\cite{8979252_liu_covar_ippm} & Covariance-aware weighted PPM in cooperative settings & TOA-only & No support for TDOA; lacks SNR or geometry-aware weighting; assumes tight anchor-UE synchronization; moderate accuracy \\
\hline
Dureppagari et al.~\cite{10632801_harish_dyspan} & Reformulated IPPM for TDOA with a 3GPP-compliant CRLB framework under multipath & TOA \& TDOA & No weighting strategy; moderate accuracy; inferior to or best matches with NLS \\
\hline
\textbf{This Work} & Two-stage refinement with high accuracy, low complexity, robustness under high GDOP and low SNR, near-optimal performance at high SNR, and 3GPP-compliant evaluation & TOA \& TDOA &  --- \\
\hline
\end{tabular}
\end{table*}
Localization algorithms have been extensively studied in the literature, with numerous efforts dedicated to developing low-complexity methods that enhance accuracy while maintaining computational efficiency. Specifically, the work in~\cite{ZHUO2024} introduced a low-complexity optimization technique for localization using successive greedy grid search and successive cancellation-list grid search to address a measurement-to-target association problem, while~\cite{5464262} provides a framework for a low-complexity least-squares algorithm that demonstrates improved accuracy; however, their performance remains comparable to NLS and falls short of more advanced approaches such as WNLS and optimization-based methods. Furthermore, in ~\cite{8947090_liu_weight_ippm, 8979252_liu_covar_ippm}, the authors introduced approaches for weighted PPM to enable enhanced cooperative localization, where the weights are determined using the gradient descent technique and the covariance of TOA measurements, respectively. While these methods demonstrated promising results for TOA-based localization, they cannot be directly applied to TDOA-based scenarios. Further, they cannot be used if the TOA values are pseudorange measurements, meaning they inherently include clock offsets and bias terms. In scenarios where there is no prior information on transmission time and bias offsets, such as clock offsets and NLOS bias, TOA values cannot be directly treated as noisy range measurements. \chb{Addressing these challenges, in our earlier work~\cite{10632801_harish_dyspan}, we proposed a reformulated version of IPPM to extend its applicability to TDOA-based localization. Additionally, we introduced a 3GPP-compliant analytical framework using the CRLB tool that accounts for 5G NR physical layer parameters and multipath fading effects, which we extend in this article for evaluating the proposed TS-WPM.} However, it is not straightforward to introduce suitable weighting mechanisms in PPMs to prevent overemphasizing or under-emphasizing specific TDOA measurements, particularly when certain anchor-to-UE links experience low SNR~\cite{dureppagari2025icc}. To overcome these limitations and achieve a balance between improved accuracy and computational feasibility, the current paper introduces a novel TS-WPM for robust cooperative localization. The proposed approach significantly enhances positioning accuracy, outperforming state-of-the-art approaches, including IPPM, NLS, and WNLS, while maintaining efficiency suitable for practical deployment. \chb{To clearly highlight the current research landscape and position our contribution, we present a comparative summary in Table~\ref{tab:related_work_summary}, which outlines the scope, strengths, and limitations of representative prior works in relation to TS-WPM.}

\subsection{Contributions}\label{sec:contributions}
The main contributions of this paper are summarized as follows.
\begin{itemize}[wide, labelwidth=!, labelindent=0pt]
    \item \textbf{{\em Two-Stage Weighted Projection Method (TS-WPM):}} We propose a novel {\em TS-WPM} for TDOA-based localization, employing a two-stage iterative refinement process. In this approach, the range measurements associated with a reference anchor are iteratively estimated and leveraged to refine the location estimate and vice versa. This iterative refinement effectively mitigates residual errors, improving positioning accuracy, particularly under challenging conditions such as low SNR and high GDOP and equal performance everywhere else. Additionally, we extend TS-WPM to support {\em cooperative localization}, enabling target UEs to utilize internode measurements for accurate localization, especially when an inadequate number of anchors (such as 2) are visible.
    \item \textbf{{\em Optimality and Robustness of TS-WPM:}} We perform a thorough derivation of the {\em error covariance} of the proposed TS-WPM and establish its optimality in high SNR regimes. Furthermore, we compute the mean squared error (MSE) for both WNLS and TS-WPM, providing analytical proof that validates the superiority of TS-WPM over WNLS in non-ideal conditions, such as high GDOP and low SNR while demonstrating identical performance under favorable conditions, thereby highlighting the robustness of TS-WPM across diverse scenarios.
    \item \textbf{{\em 3GPP-Compliant Analytical Framework and NLOS Bias:}} To rigorously evaluate the proposed approach, we develop a {\em 3GPP-compliant system-level analytical framework} using CRLB as the principal tool. Specifically, we derive the CRLB for TOA, TDOA, and positioning error while accounting for the 5G physical layer aspects~\cite{3gpp::38211}, as well as large-scale~\cite{3gpp::36814} and small-scale fading effects~\cite{3gpp::38901}. Using the developed framework, we introduce a {\em novel NLOS bias modeling} approach that incorporates underlying propagation conditions and SNR. The derived TOA error models, along with the proposed NLOS bias model, are utilized to conduct extensive Monte Carlo simulations, enabling a more realistic performance evaluation.
\end{itemize}  

\section{Novel Two-Stage Weighted Projection Method}\label{sec:novel_ts_wpm}
In this section, we introduce 1) the novel low-complexity TS-WPM, designed for improved accuracy, especially in challenging conditions such as low SNR or high GDOP, and 2) an extension of its framework to support cooperative localization, leveraging collaborative measurements to enhance accuracy in scenarios with limited visible anchors.

\subsection{Novel TS-WPM: TDOA Localization} \label{sec:novel_ts_wpm_tdoa}
To achieve a balance between improved accuracy and maintaining computational efficiency, we introduce a novel TS-WPM designed for TDOA-based localization. This new approach is particularly crucial for addressing the challenges associated with conditions such as high GDOP or low SNR regimes. In this approach, we introduce additional variables to the estimation along with the target UE location. Specifically, we estimate the range or TOA between the reference anchor and the target UE, iteratively updating this estimate using the latest UE location. This updated range estimate is then utilized to refine the UE location estimate further.

\begin{chbgrp}
Consider a network of \(B\) anchors assisting a target UE to obtain pseudorange or TOA measurements. 
The pseudorange measurements between an anchor \(b\) and a target UE, denoted by \(\hat{r}_b\), are generally defined as
\begin{align}
    \hat{r}_b = \| \boldsymbol{\theta} - \mathbf{a}_b \| + \Delta_b + v_b, \quad b \in \{1,\cdots, B\}.\label{eq:noisy_toa}
\end{align}
where \(\mathbf{a}_b\) denotes known location coordinates of the anchor \(b\), \(\boldsymbol{\theta}\) is the true UE location, \( \| \boldsymbol{\theta} - \mathbf{a}_b \| \) is the range between the anchor \(b\) and the UE, \(\Delta_b\) is the bias induced by multipath, and \(n_b\) is additive white Gaussian noise (AWGN). Assuming the pseudorange measurements \(\hat{r}_b\) are statistically independent across anchors, the covariance matrix of the TOA measurement vector \(\hat{\mathbf{r}} = [\hat{r}_1,\cdots,\hat{r}_B]^T\) is diagonal and takes the form \(\hat{\boldsymbol{\Sigma}} = \mathrm{diag}(\sigma_1^2,\cdots,\sigma_B^2)\), where \(\sigma_b^2\) is the variance of \(v_b\).

Now, without loss of generality, we assume anchor \(1\) as the reference anchor. The TDOA measurements are then defined as
\begin{align}
    \tilde{r}_b = \hat{r}_b - \hat{r}_1,\quad b \in \{2,\dots, B\},
    \label{eq:noisy_tdoa}
\end{align}

Typically, the anchor with the strongest SNR is selected as the reference anchor~\cite{buehrer2019handbook}. This choice minimizes the variance of the common reference term \(\hat{r}_1\), thereby reducing the overall correlation among TDOA measurements and improving localization robustness. Unlike pseudorange measurements, TDOA measurements share the common reference term \(\hat{r}_1\), making them mutually dependent. As a result, the covariance matrix of the TDOA measurement vector \( \tilde{\mathbf{r}} = [\tilde{r}_2, \ldots, \tilde{r}_B]^T \) becomes non-diagonal. It takes the form \(\tilde{\boldsymbol{\Sigma}} = \mathrm{diag}([\sigma_2^2, \dots, \sigma_B^2]) + \sigma_1^2 \mathbf{u}\mathbf{u}^T\), where \(\mathbf{u}\) is a \((B-1) \times 1\) vector of ones, capturing the correlation induced by the common reference measurement \(\hat{r}_1\). \end{chbgrp}Denoting an estimate of the UE location by \(\hat{\boldsymbol{\theta}}\), as part of TDOA-localization, we aim to minimize the TDOA residual error calculated as
\begin{align}
    \mathbf{R} \leftarrow \frac{1}{B-1} \sum_{b=2}^{B} \left(\tilde{r}_{b} - \tilde{r}_{b,e}(\hat{\boldsymbol{\theta}})\right)^2,
\end{align}
where \(\tilde{r}_{b,e}(\hat{\boldsymbol{\theta}})\) represents an estimate of TDOA associated with the anchor \(b\) calculated as 
\begin{align}
    \tilde{r}_{b,e}(\hat{\boldsymbol{\theta}}) = \|\hat{\boldsymbol{\theta}}-\mathbf{a}_b\| - \|\hat{\boldsymbol{\theta}}-\mathbf{a}_1\|.
    \label{eq:tdoa_est_method1}
\end{align}

\chb{In our proposed approach, we introduce an additional unknown parameter \(r_1\) in the estimation process alongside \(\boldsymbol{\theta}\), where \(r_1\) enables an accurate estimation of the range between the reference anchor and the UE, thereby reducing TDOA residual errors and improving the localization accuracy. To facilitate this, we define a new variable
\begin{align}
    \tilde{r}'_b = \hat{r}_b - \hat{r}_1^e, \quad b \in \{1, 2, \cdots, B\},
    \label{eq:revised_noisy_tdoa}
\end{align}
where \(\hat{r}_1^{\,e}\) denotes an estimate of true range from the reference anchor to the UE. Unlike the pseudorange \(\hat{r}_1\), which includes unknown clock biases and is inherently noisy, \(\hat{r}_1^{\,e}\) is iteratively refined to approach the true (bias-free) range.} Importantly, we are not modifying the actual TDOA measurements but rather introducing a new variable that serves as the basis for the proposed approach and the corresponding analytical study. The proposed method still uses conventional TDOA measurements to update UE location, residual error, and range estimates, but the analyses and refinement process are guided by \(\tilde{r}'_b\). Given \(\hat{r}_1^e\), \(\tilde{r}'_b\) become independent across anchors resembling TOA-like measurements. This ensures that, given \(\hat{r}_1^e\), \({\rm var}(\tilde{r}'_b)\) is essentially equal to \({\rm var}(\hat{r}_b)\). To incorporate this refinement into the estimation process, we redefine the residual error to be minimized as  
\begin{align}
    \mathbf{R}' \leftarrow \frac{1}{B} \sum_{b=1}^{B} \left(\tilde{r}_{b} - \tilde{r}'_{b,e}(\hat{\boldsymbol{\theta}},\hat{r}_1^e)\right)^2,
\end{align}  
where \(\tilde{r}_1 = 0\) for the reference anchor (\(b=1\)), and \(\tilde{r}'_{b,e}(\hat{\boldsymbol{\theta}},\hat{r}_1^e)\) represents the refined TDOA estimate, computed as  
\begin{align}
    \tilde{r}'_{b,e}(\hat{\boldsymbol{\theta}},\hat{r}_1^e) = \|\hat{\boldsymbol{\theta}}-\mathbf{a}_b\| - \hat{r}_1^e.
    \label{eq:tdoa_est_method2}
\end{align}

We now discuss the significance of this new formulation and its advantages in simplifying the process of generating weights by avoiding the estimation of the full covariance matrix, as required in methods like WNLS~\cite{buehrer2019handbook}. As mentioned earlier, with \(\hat{r}_1^e\) available, the measurements described in~\eqref{eq:revised_noisy_tdoa} become independent of each other. As a result, the covariance matrix becomes a diagonal matrix, where the diagonal elements represent the variances of the different TOA equivalent measurements, as opposed to the non-diagonal covariance matrix we observe for the conventional TDOA measurements. Therefore, the new covariance matrix, denoted by \( \tilde{\boldsymbol{\Sigma}} \), takes the form
\begin{align}
\tilde{\boldsymbol{\Sigma}} = {\rm diag}({\rm var}(\tilde{r}'_1),\cdots,{\rm var}(\tilde{r}'_B)) = {\rm diag}({\rm var}(\hat{r}_1),\cdots,{\rm var}(\hat{r}_B)).
    \label{eq:revised_tdoa_covariance}
\end{align}

This newly defined covariance matrix \(\tilde{\boldsymbol{\Sigma}}\) is then utilized to generate weights for the proposed TS-WPM. A key advantage of this approach is the significant reduction in computational complexity, as it eliminates the need to estimate and invert the full covariance matrix. Since \(\tilde{\boldsymbol{\Sigma}}\) is a diagonal matrix, inverting it simply involves inverting the individual diagonal elements, which is computationally efficient. The weights are generated as
\begin{align}
    W_b = \frac{\frac{1}{{\rm var}(\hat{r}_b)}}{\sum_{i=1}^{B} \frac{1}{{\rm var}(\hat{r}_i)}}, \quad b = {1, 2, \cdots, B}.
    \label{eq:weights_novel_ppm}
\end{align}
\begin{algorithm}[htbp]
\small
\captionsetup{font=small}
\caption{Novel TS-WPM for TDOA Localization}
\label{alg:novel_ts_wpm_tdoa}
\KwData{Initial UE estimate $\hat{\boldsymbol{\theta}}^0$, Initial range estimate $\hat{r}_1^{e,0}$, Convergence threshold $\epsilon$, Maximum consecutive iterations to terminate $l$, Weights $W_b$}
\KwResult{Final estimate $\hat{\boldsymbol{\theta}}$}
\SetAlgoNlRelativeSize{0}
\SetNlSty{}{}{)}
\DontPrintSemicolon

$\mathbf{R}^0 \leftarrow \frac{1}{B} \sum_{b=1}^{B} \left(\tilde{r}_b - \big(\|\hat{\boldsymbol{\theta}}^0-\mathbf{a}_b\| - \hat{r}_1^{e,0}\big)\right)^2$\;
$k \leftarrow 1$; $c \leftarrow 0$\;
\While{True}{    
    $\hat{\boldsymbol{\theta}}^k \leftarrow \sum_{b=1}^{B} W_b\left(\mathbf{a}_b + \left(\tilde{r}_b + \hat{r}_1^{e,k-1}\right)\frac{\hat{\boldsymbol{\theta}}^{k-1}-\mathbf{a}_b}{\|\hat{\boldsymbol{\theta}}^{k-1}-\mathbf{a}_b\|}\right)$\;\label{step:novel_ts_wpm_tdoa_loc_update}
    
    $\mathbf{R}^k \leftarrow \frac{1}{B} \sum_{b=1}^{B} \left(\tilde{r}_b - \left(\|\hat{\boldsymbol{\theta}}^k-\mathbf{a}_b\| - \hat{r}_1^{e,k-1}\right)\right)^2$\; \label{step:novel_ts_wpm_tdoa_res_err}
    
    \If{$|\mathbf{R}^k-\mathbf{R}^{k-1}| < \epsilon$}{
        $c \leftarrow c + 1$\;
        \If{$c \geq l$}{
            $\hat{\boldsymbol{\theta}} \leftarrow \hat{\boldsymbol{\theta}}^k$\;
            break\;
        }
    }
    \Else{
        $c \leftarrow 0$\;
    }
    $\hat{r}_1^{e,k} \leftarrow \sum_{b=1}^{B} W_b\left(\|\hat{\boldsymbol{\theta}}^{k-1}-\mathbf{a}_b\| - \tilde{r}_b \right)$\;\label{step:novel_ts_wpm_toa_range_update}
    $k \leftarrow k + 1$\;
}
\end{algorithm}

The proposed novel TS-WPM is detailed in Algorithm~\ref{alg:novel_ts_wpm_tdoa}, where weights are applied during the updates of both the UE location estimates and the range estimates. \chb{As illustrated, the algorithm starts by initializing the UE location estimate, denoted by \(\hat{\boldsymbol{\theta}}^0\), which is usually assigned the coordinates of the nearest anchor, and the initial range estimate associated with the reference anchor, denoted by \(\hat{r}_1^{e,0}\) is set to \(\|\hat{\boldsymbol{\theta}}^0 - \mathbf{a}_1\|\) or the actual range estimate \(\hat r_1\) if operating in high SNR regime}. At the \(k\)-th iteration, the UE location estimate, \(\hat{\boldsymbol{\theta}}^k\), is iteratively updated in step~\ref{step:novel_ts_wpm_tdoa_loc_update} using the previous location estimate, \(\hat{\boldsymbol{\theta}}^{k-1}\), along with the refined range estimate, \(\hat{r}_1^{e,k-1}\), obtained from iteration \(k-1\). Concurrently, the TDOA residual error is updated in step~\ref{step:novel_ts_wpm_tdoa_res_err} using the position estimate at the \(k\)-th iteration and the range estimate from the previous iteration. This range estimate is iteratively updated and accurately estimated in step~\ref{step:novel_ts_wpm_toa_range_update}, which, in turn, refines the UE location estimate in step~\ref{step:novel_ts_wpm_tdoa_loc_update} in the following iterations. This two-stage iterative process, by effectively mitigating the impact of bias offsets, including NLOS bias, leads to a continuous enhancement of positioning accuracy. This approach is particularly advantageous in scenarios with significant bias offsets, e.g., large NLOS bias and low SNR, where traditional methods struggle to maintain accuracy. The two-stage iterative refinement in TS-WPM has a conceptual resemblance to Turbo decoding principles~\cite{539767}, where extrinsic information is exchanged between decoders to improve error correction. Similarly, in TS-WPM, refining range estimates and UE location updates iteratively improves positioning accuracy. However, unlike Turbo decoding, where exchanged information is independent in terms of redundant parity, the refined information in TS-WPM is correlated, as both updates are derived from the same set of TDOA measurements.

\subsection{Novel TS-WPM: Cooperative Localization} \label{sec:novel_wppm_coop}
In this section, we extend the novel TS-WPM approach described in Algorithm~\ref{alg:novel_ts_wpm_tdoa} to enable cooperative localization, where multiple target UEs leverage internode measurements to enhance positioning accuracy, especially when only a limited number of anchors (such as 2) are visible. In our cooperative localization framework, we assume that TDOA measurements are available between visible anchors and UEs, while TW-TOA measurements are utilized for inter-target UE measurements. In particular, modern positioning technologies such as sidelink positioning (SLP)~\cite{3gpp::38355} enable a wide range of measurement techniques, including TOA, angle of arrival (AOA), round--trip time (RTT), and TDOA, between target UEs. However, we specifically adopt TW-TOA for internode measurements for the following key reasons: 1) TW-TOA eliminates the need for precise synchronization among collaborating nodes~\cite{buehrer2019handbook}. 2) Many existing positioning technologies, such as ultra-wideband (UWB)~\cite{9810941} and WiFi-based localization~\cite{ibrahim2018verification}, predominantly rely on TW-TOA measurements for ranging. By utilizing TW-TOA for cooperative localization, our approach remains flexible and interoperable with different positioning technologies, facilitating seamless integration across various deployment scenarios.  3) The geometric configuration of collaborating target UEs is often suboptimal, particularly in indoor environments or clustered deployments. Relying on relative measurements such as TDOA in these cases can lead to significant localization errors due to high GDOP. By leveraging TW-TOA measurements instead, we improve localization robustness, especially in non-ideal geometries.

The extended framework for cooperative localization is outlined in Algorithm~\ref{alg:novel_ts_wpm_coop}. As part of this, we aim to jointly localize all target UEs, denoted by \(N_u\). For each target UE \(n\), we define a set of cooperative nodes \(S_n\), with \(|S_n|\) denoting its cardinality. Furthermore, for each target UE \(n\), we begin by initializing location estimate \(\hat{\boldsymbol{\theta}}_n^0\), the reference anchor range estimate \(\tilde{r}_{1,n}^{e,0}\), and a flag \(F_n\) indicating whether the target UE is localized (\(F_n=1\)) or not (\(F_n=0\)), i.e., when the iterative loop converges for the target UE \(n\). In this formulation, \(\tilde{r}_{b,n}\) represents TDOA measurements between the anchor \(b\) and the target UE \(n\), while \(\hat{r}_{u,n}\) denotes TOA measurements between two cooperating UEs \(u\) and \(n\). Next, we iteratively refine the location estimates, residual errors, and the range measurements associated with the respective reference anchors. In step~\ref{step:novel_ts_wpm_coop_loc_update1} of the algorithm, each target UE location is updated based on TDOA measurements from visible anchors, employing weights \(W_b\). Step~\ref{step:novel_ts_wpm_coop_loc_update2} incorporates cooperative measurements between UEs in \(S_n\), applying weights \(W_u\). The weights \(W_u\) are generated similar to \(W_b\) as given in~\eqref{eq:weights_novel_ppm}. Residual errors are then updated in steps~\ref{step:novel_ts_wpm_coop_res_err1} and~\ref{step:novel_ts_wpm_coop_res_err2}. At convergence, the final estimate for the UE is updated, and \(F_n\) is set to 1, indicating convergence. This process continues for all UEs within the loop in step~\ref{step:for_loop_all_ues}, terminating once every \(F_n\) equals 1 (that is, all target UEs have been localized). Throughout this procedure, the range estimates \(\hat{r}_{1,n}^{e,k}\) in step~\ref{step:novel_ts_wpm_coop_range_update} are iteratively refined, further enhancing location accuracy.
\begin{algorithm}[t]
\small
\captionsetup{font=small}
\caption{Novel TS-WPM for Cooperative Localization}
\label{alg:novel_ts_wpm_coop}
\KwData{Number of UEs \(N_u\), Initial UE estimates \(\left[\hat{\boldsymbol{\theta}}_1^0,\,\hat{\boldsymbol{\theta}}_2^0,\,\cdots, \hat{\boldsymbol{\theta}}_{N_u}^0\right]\), Initial range estimates for the corresponding reference anchors \(\left[\hat{r}_{1,1}^{e,0},\,\tilde{r}_{1,2}^{e,0},\cdots\tilde{r}_{1,N_u}^{e,0}\right]\), Convergence threshold \(\epsilon\), Maximum consecutive iterations to terminate \(l\), Weights \(W_b\), Weights \(W_u\), Initialize \(F_n=0, n = 1,2,\cdots,N_u\)}
\KwResult{Final estimate $\hat{\boldsymbol{\theta}}$}
\SetAlgoNlRelativeSize{0}
\SetNlSty{}{}{)}
\DontPrintSemicolon

$k \leftarrow 1$\;
\For{$n = 1$ \KwTo $N_u$}{
    $\mathbf{R}_n^0 \leftarrow \frac{1}{B} \sum_{b=1}^{B} \left(\hat{r}_{b,n} - \left(\|\hat{\boldsymbol{\theta}}_n^0-\mathbf{a}_b\| - \tilde{r}_{1,n}^0\right)\right)^2$\;
    $\mathbf{R}_n^0 \leftarrow \mathbf{R}_n^0 + \frac{1}{|S_n|} \sum_{u\in S_n}
    \left(\hat{r}_{u,n} - \left(\|\hat{\boldsymbol{\theta}}_n^0-\hat{\boldsymbol{\theta}}_u^0\|\right)\right)^2$\;
    }
\For{$n = 1$ \KwTo $N_u$}{\label{step:for_loop_all_ues}
\While{any $F_n = 0$ for $n = 1$ to $N_u$}{
    $\hat{\boldsymbol{\theta}}_n^k \leftarrow \frac{1}{B} \sum_{b=1}^{B} W_b\left(\mathbf{a}_b + \left(\tilde{r}_{b,n} + \hat{r}_{1,n}^{e,k-1}\right)\frac{\hat{\boldsymbol{\theta}}_n^{k-1}-\mathbf{a}_b}{\|\hat{\boldsymbol{\theta}}_n^{k-1}-\mathbf{a}_b\|}\right)$\;\label{step:novel_ts_wpm_coop_loc_update1}

    $\hat{\boldsymbol{\theta}}_n^k \leftarrow \hat{\boldsymbol{\theta}}_n^k + \frac{1}{|S_n|} \sum_{u\in S_n} W_u\left(\hat{\boldsymbol{\theta}}_u^{k-1} + \hat{r}_{u,n}\frac{\hat{\boldsymbol{\theta}}^{k-1}-\hat{\boldsymbol{\theta}}_u^{k-1}}{\|\hat{\boldsymbol{\theta}}^{k-1}-\hat{\boldsymbol{\theta}}_u^{k-1}\|}\right)$\;\label{step:novel_ts_wpm_coop_loc_update2}
    
    $\mathbf{R}_n^k \leftarrow \frac{1}{B} \sum_{b=1}^{B} \left(\tilde{r}_{b,n} - \left(\|\hat{\boldsymbol{\theta}}_n^k-\mathbf{a}_b\| - \hat{r}_{1,n}^{e,k-1}\right)\right)^2$\; \label{step:novel_ts_wpm_coop_res_err1}

    $\mathbf{R}_n^k \leftarrow \mathbf{R}_n^k + \frac{1}{|S_n|} \sum_{u\in S_n} \left(\hat{r}_{u,n} - \left(\|\hat{\boldsymbol{\theta}}_n^k-\hat{\boldsymbol{\theta}}_u^{k-1}\|\right)\right)^2$\; \label{step:novel_ts_wpm_coop_res_err2}
    
    \If{$|\mathbf{R}_n^k-\mathbf{R}_n^{k-1}| < \epsilon$}{
        $C_n \leftarrow C_n + 1$\;
        \If{$c \geq l$}{
            $\hat{\boldsymbol{\theta}}_n \leftarrow \hat{\boldsymbol{\theta}}_n^k$\;
            break\;
        }
    }
    \Else{
        $c \leftarrow 0$\;
    }
    $\hat{r}_{1,n}^{e,k} \leftarrow \frac{1}{B} \sum_{b=1}^{B} W_b\left(\|\hat{\boldsymbol{\theta}}_n^{k-1}-\mathbf{a}_b\| - \tilde{r}_{b,n} \right)$\;\label{step:novel_ts_wpm_coop_range_update}
    $k \leftarrow k + 1$\;
}
}
\end{algorithm}

\section{Optimality and Robustness of TS-WPM: Analytical Study}\label{sec:analytical_study_tswpm}
In this section, we present a comprehensive analytical proof demonstrating the conditions under which the proposed two-stage approach is optimal and when it outperforms WNLS. To achieve this, we first derive the error covariance matrix for the novel TS-WPM by using the alternative formulation outlined in~\eqref{eq:revised_noisy_tdoa} as basis. We then compare it with the error covariance for the two-stage estimation employing the maximum likelihood estimation (MLE) as described in~\cite{murphy_topel} and discuss the optimality of the proposed TS-WPM. Afterward, we derive the error covariance employing WNLS for the conventional TDOA formulation given in~\eqref{eq:noisy_tdoa}. Subsequently, we derive and compare the MSE for both methods, TS-WPM and WNLS, discussing the conditions under which TS-WPM surpasses WNLS. Additionally, we provide an alternative mathematical intuition to elucidate why TS-WPM outperforms WNLS by comparing the normalization factors used in both methods alongside the enhanced accuracy that the two-stage approach offers. As a foundation for this analysis, we begin by introducing key preliminaries, including a discussion of conventional and alternative TDOA formulations and the derivation of the Fisher Information Matrix (FIM) for both methods, which we will use later in the analysis.

\subsection{Key Preliminaries}\label{sec:fim_conv_reform_tdoa}
{\em Method 1: Conventional TDOA Formulation. } The conventional TDOA measurements are as given in~\eqref{eq:noisy_tdoa}. Defining the TDOA measurement vector as \(\tilde{\mathbf{r}} = [\tilde{r}_2, \dots, \tilde{r}_B]^T\) and the variance of \(\hat{r}_b\) as \(\sigma_b^2\), the covariance matrix of \(\tilde{\mathbf{r}}\) is given by
\begin{align}
  \mathbf{C}_1 = \mathrm{diag}([\sigma_2^2, \dots, \sigma_B^2]) + \sigma_1^2 \mathbf{u}\mathbf{u}^T,
  \label{eq:cov_tdoa_scen1}
\end{align}  
where \(\mathbf{u}\) is a \((B-1) \times 1\) vector of ones, capturing the correlation induced by the common reference measurement \(\hat{r}_1\). Assuming 2D estimation, we define the parameter vector with UE coordinates as \(\boldsymbol{\theta}=[x,y]\). The Jacobian matrix of \(\tilde{\mathbf{r}}\) with respect to \(\boldsymbol{\theta}\) is expressed as  
\begin{align}
  \mathbf{H}_1 = \begin{bmatrix}
  \frac{x - x_2}{d_2} - \frac{x - x_1}{d_1} & \frac{y - y_2}{d_2} - \frac{y - y_1}{d_1} \\
  \vdots & \vdots \\
  \frac{x - x_B}{d_B} - \frac{x - x_1}{d_1} & \frac{y - y_B}{d_B} - \frac{y - y_1}{d_1}
  \end{bmatrix}.
  \label{eq:jacobian_tdoa}
\end{align}  

With this formulation, the FIM for \(\hat{\mathbf{r}}\) is expressed as  
\begin{align}
  \mathbf{F}_1 = \mathbf{H}_1^T \mathbf{C}_1^{-1} \mathbf{H}_1.
  \label{eq:fim_tdoa_scen1}
\end{align}

{\em Method 2: Alternative TDOA Formulation.} As outlined in Section~\ref{sec:novel_ts_wpm_tdoa}, we introduce an additional parameter \(r_1\) to the parameter vector. Considering 2D estimation, the parameter vector is defined as \(\boldsymbol{\theta}_2 = [\boldsymbol{\theta},r_1] = [x,y,r_1]\), where \(r_1\) facilities an accurate range estimation associated with the reference anchor. To incorporate the estimate of \(r_1\), we define an alternative TDOA formulation as given in~\eqref{eq:revised_noisy_tdoa}, and the corresponding covariance matrix is of the form  
\begin{align}
  \mathbf{C}_2 = \mathrm{diag}([\sigma_1^2, \dots, \sigma_B^2]).
  \label{eq:cov_tdoa_scen2}
\end{align}  

Denoting a vector of alternative TDOA measurements by \(\tilde{\mathbf{r}}'\), the Jacobian matrix of \(\tilde{\mathbf{r}}'\) with respect to \(\boldsymbol{\theta}_2\) is then defined as
\begin{align}
    \mathbf{H}_2 = [\mathbf{H}_t\, \mathbf{v}],
    \label{eq:jacobian_pseudo_toa}
\end{align}
where \(\mathbf{H}_t\) is given by  
\begin{align}
    \mathbf{H}_t = \begin{bmatrix}
    \frac{x - x_1}{d_1} \cdots \frac{x - x_B}{d_B}\\
      \frac{y - y_1}{d_1} \cdots \frac{y - y_B}{d_B}
    \end{bmatrix}^T,
    \label{eq:jacobian_toa}
\end{align}  
and \(\mathbf{v}\) is a \(B\times 1\) vector of ones.

The FIM for \(\tilde{\mathbf{r}}\) is given by
\begin{align}
  \mathbf{F}_2 = \mathbf{H}_2^T \mathbf{C}_2^{-1} \mathbf{H}_2.
  \label{eq:fim_tdoa_scen2}
\end{align}

Expanding \(\mathbf{F}_2\) in~\eqref{eq:fim_tdoa_scen2}, we obtain
\begin{align}
    \mathbf{F}_2 = \begin{bmatrix}
    \mathbf{H}_t^T\mathbf{C}_2^{-1}\mathbf{H}_t & -\mathbf{H}_t^T\mathbf{C}_2^{-1}\mathbf{v}\\
    -\mathbf{v}^T\mathbf{C}_2^{-1}\mathbf{H}_t & \mathbf{v}^T\mathbf{C}_2^{-1}\mathbf{v}
    \end{bmatrix}.
\end{align}  

Applying Schur complement~\cite{MoeYuanP12010}, we derive the effective FIM for the position parameter \(\boldsymbol{\theta}=[x,y]\) as
\begin{align}
    \mathbf{F}_{2,[x,y]}=\mathbf{H}_t^T\mathbf{C}_2^{-1}\mathbf{H}_t - \mathbf{H}_t^T\mathbf{C}_2^{-1}\mathbf{v}\left(\mathbf{v}^T\mathbf{C}_2^{-1}\mathbf{v}\right)^{-1}\mathbf{v}^T\mathbf{C}_2^{-1}\mathbf{H}_t.
    \label{eq:efim_tdoa_scen2}
\end{align}

\chb{The second term in~\eqref{eq:efim_tdoa_scen2} quantifies the loss of Fisher information resulting from the estimation of the additional parameter \(r_1\). When additional (nuisance) parameters are jointly estimated along with the parameters of interest (in this case, UE position), the effective information available for position estimation is reduced. The term~\eqref{eq:efim_tdoa_scen2} explicitly captures the amount of information ``spent'' on estimating \(r_1\), thereby quantifying the degradation of localization performance compared to a scenario where \(r_1\) is known a priori.} Not surprisingly, both formulations in~\eqref{eq:noisy_tdoa} and~\eqref{eq:revised_noisy_tdoa} contain the same amount of information, i.e., \(\mathbf{F}_1=\mathbf{F}_{2,[x,y]}\), resulting in identical CRLB.

\subsection{Optimality and Robustness of TS-WPM}\label{sec:optimality_robustness_ts_wpm}
We now apply the proposed TS-WPM for Method 2, which uses a two-stage iterative refinement approach, as outlined in Algorithm~\ref{alg:novel_ts_wpm_tdoa}. Notably, TS-WPM is designed to iteratively minimize residual errors, ensuring convergence, which is also observed in methods like WNLS. Assuming convergence, the final estimate can be approximated as
\begin{align}
\hat{\boldsymbol{\theta}}_{\mathrm{tswpm}} = \boldsymbol{\theta} + (\mathbf{H}_2(\boldsymbol{\theta}))^T\mathbf{W}_2\big(\tilde{\mathbf{r}} - \tilde{\mathbf{r}}'_e(\boldsymbol{\theta},r_1)\big),
\label{eq:nwpm_final_est}
\end{align}
where \( \boldsymbol{\theta} \) is the true position parameter, \(\mathbf{H}_2(\boldsymbol{\theta})\) is the Jacobian matrix of \(\tilde{\mathbf{r}}\) with respect to \(\boldsymbol{\theta}\) and is equal to \(\mathbf{H}_2\) as given in~\eqref{eq:jacobian_pseudo_toa}, \(\tilde{\mathbf{r}}'_e(\boldsymbol{\theta},r_1)\) is a vector of TDOA estimates as a function of \(\boldsymbol{\theta}\) and \(r_1\) as given in~\eqref{eq:tdoa_est_method2}, and the weight matrix \(\mathbf{W}_2\) is defined as \(\mathbf{W}_2 = \frac{\mathbf{C}_2^{-1}}{{\rm trace}(\mathbf{C}_2^{-1})}\), where \(\frac{1}{{\rm trace}(\mathbf{C}_2^{-1})}\) serves as a normalization factor to prevent overemphasizing or under-emphasizing certain measurements. We now derive the error covariance and MSE of TS-WPM.
\begin{lemma}[Error Covariance and MSE of TS-WPM]\label{lemma::cov::mse::tswpm}
The error covariance of TS-WPM, denoted by \({\rm cov}(\hat{\boldsymbol{\theta}}_{\mathrm{tswpm}}-\boldsymbol{\theta})\), and its MSE, denoted by \({\rm MSE}_{\mathrm{tswpm}}\), can be obtained as:
\begin{align}
    {\rm cov}(\hat{\boldsymbol{\theta}}_{\mathrm{tswpm}}-\boldsymbol{\theta}) = & \frac{1}{\left({\rm trace}(\mathbf{C}_2^{-1})\right)^2}\left(\mathbf{H}_2^T\mathbf{C}_2^{-1}\mathbf{H}_2\right),\label{eq:cov_tswpm_main}\\
    {\rm MSE}_{\mathrm{tswpm}} = & \frac{1}{\left({\rm trace}(\mathbf{C}_2^{-1})\right)^2}\sum_{i=1}^K\lambda_i\left(\mathbf{F}_{2,[x,y]}\right),  
    \label{eq:mse_tswpm_main}
\end{align}
where \(\mathbf{C}_2\), \(\mathbf{H}_2\), and \(\mathbf{F}_{2,[x,y]}\) are as given in~\eqref{eq:cov_tdoa_scen2}, ~\eqref{eq:jacobian_pseudo_toa}, and~\eqref{eq:efim_tdoa_scen2}, respectively, and \(\lambda_i\left(\mathbf{F}_{2,[x,y]}\right)\) denotes the eigenvalues of \(\mathbf{F}_{2,[x,y]}\).
\end{lemma}
\begin{IEEEproof} 
See Appendix~\ref{app::cov::mse::tswpm}.
\end{IEEEproof}
This result will be further used in subsequent analyses discussing the optimality and robustness of TS-WPM.

We now establish that TS-WPM approaches MLE in a high SNR regime.
\begin{theorem}[Optimality of TS-WPM]\label{thm::optimality::tswpm}
Consider a two-stage estimation framework where the first-stage estimate refines the second-stage estimate, similar to the approach described in Algorithm~\ref{alg:novel_ts_wpm_tdoa}, where the reference anchor range is accurately estimated in the first stage, which enhances the UE location estimate in the second stage. Suppose \(\sigma_1^2 \to 0\), i.e., the SNR of the reference anchor tends to infinity. Then, 
\begin{align}
    \mathrm{cov}\bigl(\hat{\boldsymbol{\theta}}_{\mathrm{tswpm}} - \boldsymbol{\theta}\bigr)
    ~\longrightarrow~
    \boldsymbol{\Sigma}_{\boldsymbol{\theta}}^{\mathrm{MLE}},
\end{align}
where \(\boldsymbol{\Sigma}_{\boldsymbol{\theta}}^{\mathrm{MLE}}\) is the error covariance of MLE. In other words, TS-WPM is asymptotically optimal in a high SNR regime.
\end{theorem}
\begin{IEEEproof} 
See Appendix~\ref{app::optimality::tswpm}.
\end{IEEEproof}

To compare the performance of TS-WPM and WNLS, we derive the error covariance and MSE of WNLS by applying WNLS to Method 1.
\begin{lemma}[Error Covariance and MSE of WNLS]\label{lemma::cov::mse::wnls}
The error covariance of WNLS, denoted by \({\rm cov}(\hat{\boldsymbol{\theta}}_{\mathrm{wnls}}-\boldsymbol{\theta})\), and its MSE, denoted by \({\rm MSE}_{\mathrm{wnls}}\), are given by:
\begin{align}
    {\rm cov}(\hat{\boldsymbol{\theta}}_{\mathrm{wnls}}-\boldsymbol{\theta}) =& \left(\mathbf{H}_1^T\mathbf{C}_1^{-1}\mathbf{H}_1\right)^{-1},\label{eq:cov_wnls_main}\\
    {\rm MSE}_{\mathrm{wnls}} = & \sum_{i=1}^K \frac{1}{\lambda_i\left(\mathbf{H}_1^T\mathbf{C}_1^{-1}\mathbf{H}_1\right)},  
    \label{eq:mse_wnls_main}
\end{align}
where \(\mathbf{C}_1\), \(\mathbf{H}_1\) are as given in~\eqref{eq:cov_tdoa_scen1} and ~\eqref{eq:jacobian_tdoa}, respectively, and \(\lambda_i\left(\mathbf{H}_1^T\mathbf{C}_1^{-1}\mathbf{H}_1\right)\) denotes the eigenvalues of \(\mathbf{H}_1^T\mathbf{C}_1^{-1}\mathbf{H}_1\).
\end{lemma}
\begin{IEEEproof} 
See Appendix~\ref{app::cov::mse::wnls}.
\end{IEEEproof}

Using Lemma~\ref{lemma::cov::mse::tswpm} and Lemma~\ref{lemma::cov::mse::wnls}, we now establish the robustness of TS-WPM.
\begin{prop}[Robustness of TS-WPM]\label{prop::robustness::tswpm}
Under non-ideal conditions such as high GDOP scenarios, the MSE of TS-WPM and WNLS can be approximated as
\begin{align}
    \mathrm{MSE}_{\mathrm{tswpm}} &\approx \frac{1}{\sum_{b=1}^B \frac{1}{\sigma_b^2}}, 
    \label{eq:mse_tswpm_approx}\\
    \mathrm{MSE}_{\mathrm{wnls}} &\approx \frac{1}{\lambda_{\min}\bigl(\mathbf{H}_1^T\mathbf{C}_1^{-1}\mathbf{H}_1\bigr)},
    \label{eq:mse_wnls_approx}
\end{align}
where $\sigma_b^2$ is the variance of the alternative TDOA measurement $\tilde{r}_b'$ in \eqref{eq:revised_noisy_tdoa}, and $\lambda_{\min}\bigl(\mathbf{H}_1^T\mathbf{C}_1^{-1}\mathbf{H}_1\bigr)$ denotes the smallest eigenvalue of $\mathbf{H}_1^T\mathbf{C}_1^{-1}\mathbf{H}_1$. Consequently, in high GDOP regimes, we have $\mathrm{MSE}_{\mathrm{tswpm}} < \mathrm{MSE}_{\mathrm{wnls}}$, demonstrating that the proposed TS-WPM outperforms WNLS under non-ideal conditions.
\end{prop}
\begin{IEEEproof}
See Appendix~\ref{app::robustness::tswpm}.
\end{IEEEproof}

\subsection{Alternative Intuition: Robustness of TS-WPM}\label{sec:alt_math_intuition}
In WNLS, we have the normalization factor taking the form \(\left(\mathbf{H}_1^T\mathbf{W}_1\mathbf{H}_1\right)^{-1}\), where \(\mathbf{W}_1\) is an inverse of the covariance matrix. If \(\mathbf{H}_1^T\mathbf{W}_1\mathbf{H}_1\) is ill-conditioned (often observed in high GDOP scenarios), direct inversion can lead to numerical instability and excessively large step sizes for certain TDOA measurements, resulting in overshooting. To understand this, consider the eigendecomposition. Since \(\mathbf{H}_1^T\mathbf{W}_1\mathbf{H}_1\) is a symmetric matrix and \(\mathbf{W}_1\) is symmetric and positive semi-definite, it can be diagonalized as
\begin{align}
    \mathbf{H}_1^T\mathbf{W}_1\mathbf{H}_1 = \mathbf{U}\boldsymbol{\Lambda} \mathbf{U}^T,
\end{align}
where \(\mathbf{U}\) is an unitary matrix and \(\boldsymbol{\Lambda}\) is a diagonal matrix containing eigenvalues \(\lambda_i\) of \(\mathbf{H}_1^T\mathbf{W}_1\mathbf{H}_1\). Then, the inverse is given by
\begin{align}
    \left(\mathbf{H}_1^T\mathbf{W}_1\mathbf{H}_1\right)^{-1} = \mathbf{U}\boldsymbol{\Lambda}^{-1} \mathbf{U}^T.
\end{align}

For an ill-conditioned matrix, \(\boldsymbol{\Lambda}^{-1}\) amplifies directions with small \(\lambda_i\), leading to large updates and overshooting. A computationally stable alternative is to normalize the location update using
\begin{align}
    \alpha = \frac{1}{\text{trace}\left(\mathbf{H}_1^T\mathbf{W}_1\mathbf{H}_1\right)},
    \label{eq:alt_normalization}
\end{align}
which scales the TDOA measurements based on the sum of eigenvalues, reducing sensitivity to small \(\lambda_i\). This approach mitigates numerical instability while maintaining effective convergence. Note that, in the proposed approach outlined in Algorithm~\ref{alg:novel_ts_wpm_tdoa} and~\eqref{eq:nwpm_final_est}, we use a variant of~\eqref{eq:alt_normalization} by using \(\frac{1}{{\rm trace}(\mathbf{C}_2^{-1})}\) as a normalization factor. This, in addition to the two-stage refinement process, results in the proposed approach providing superior performance to WNLS, particularly in challenging conditions such as high GDOP. Our evaluations also show that, in low SNR conditions under multipath propagation, the proposed TS-WPM surpasses WNLS, which we discuss in detail in Section~\ref{sec:results_discussion}.

\section{Analytical Evaluation Framework}\label{sec:analytical_eval_framework}
In this section, we develop a comprehensive analytical framework to evaluate the proposed TS-WPM, accounting for the effects of multipath propagation and AWGN. The CRLB serves as the principal tool for this framework. 
For line-of-sight (LOS) channels, the CRLB for TOA and position error has been well-documented, with accurate closed-form expressions readily available~\cite{MoeYuanP12010,4753258}. However, analysis for NLOS channels presents substantial challenges. A common approach is to incorporate a Gamma-distributed NLOS bias into the analysis~\cite{QiKo2006,6427618,dureppagari_uav1_10139944}, which lacks a functional relationship with the underlying radio propagation conditions and the physical layer aspects. 
This limitation makes this approach impractical when we consider a system model that accounts for large-scale fading, small-scale fading, and physical layer aspects, e.g., positioning reference signal (PRS) configurations. To address these shortcomings, we propose an analytical framework capable of determining the best achievable TOA error and positioning accuracy in the presence of multipath propagation. It is important to note that this framework does not aim to mitigate the effects of NLOS bias or multipath. Instead, it aims to characterize the best achievable performance under these conditions. To this end, we first derive the CRLB for TOA estimation between an anchor \(b\) and a target UE without loss of generality. We then extend this analysis to derive the CRLB for position estimates of the target UE based on TOA and TDOA measurements. Additionally, we derive the CRLB of TOA under AWGN as a special case of the multipath scenario. Furthermore, using the developed analytical framework, we propose a novel approach to model NLOS bias. This approach leverages detailed link characteristics, including physical layer parameters, large-scale fading, and small-scale fading, to accurately capture the bias induced by multipath propagation. By incorporating these elements, our approach ensures a more realistic and practical representation of NLOS effects, effectively bridging the gap between theoretical models and real-world scenarios.

\subsection{CRLB Analysis: CIR and AWGN}\label{sec:crlb_toa_cir_awgn}
In this section, we derive the CRLB for TOA error in the presence of multipath propagation. As a special case of this, we also derive the CRLB for TOA error under AWGN. Let \(\mathcal{N}_{b} = \{1, 2, \ldots, B\}\) represent the set of anchors assigned to a target UE for PRS transmission. \chb{We adopt an orthogonal frequency division multiplexing (OFDM)-based received signal model, as 5G NR employs OFDM as its primary waveform~\cite{3gpp::38211}. However, OFDM-specific impairments such as timing and frequency synchronization errors, phase noise, and high peak-to-average power ratio (PAPR) are not explicitly modeled, as they fall outside the scope of this work. This is because we assume that TOA measurements are already available, and the received signal model is used solely for deriving CRLB expressions, which in turn are used to formulate TOA error models for Monte Carlo simulations in evaluating the performance of the proposed TS-WPM algorithm.} The time domain OFDM received signal between the anchor \(b\) and the target UE can be represented in matrix form as
\begin{align}
    \boldsymbol{r}_b =  \sqrt{P_b} \boldsymbol{F}^H  \boldsymbol{X}_b  \boldsymbol{\Gamma_b}  \boldsymbol{F}_L \boldsymbol{h}_b + \boldsymbol{v},\quad b\in\mathcal{N}_{b},
    \label{eq:rxd_sgl_mat_form}
\end{align}
where $\boldsymbol{r}_b$ denotes the time-domain received PRS signal of length $N$, \(P_b\) is the received power, $\boldsymbol{v}\sim {\cal CN}({\bf 0},\sigma^2 {\bf I})$ represents AWGN, \(\boldsymbol{X}_b\) represents transmitted PRS symbols, $\boldsymbol{h}_b$ is the observed channel impulse response (CIR) between anchor \(b\) and the target UE of length \(L\), $\boldsymbol{F}_L$ is a Discrete Fourier Transform (DFT) matrix of size $N\times L$, while $\boldsymbol{F}^H$ is an inverse DFT (IDFT) matrix of size $N\times N$. Note that each delay tap of $\boldsymbol{h}_b$ translates to a phase ramp in the frequency domain, which is captured in 
\begin{align}
    \boldsymbol{\Gamma} = {\rm diag}\left(e^{-j2\pi\nabla f\tau_{d,b}(-\frac{N}{2})}\ldots e^{-j2\pi\nabla f\tau_{d,b}(\frac{N}{2}-1)}\right),\notag
\end{align}
where \(\nabla f\) denotes the sub-carrier spacing (SCS) and \(\tau_{d,b}\) is the propagation delay, determining TOA between anchor \(b\) and the target UE. 

We now present the CRLB of TOA estimation between the target UE and anchor \(b\) as follows. 
\begin{theorem}[CRLB of TOA for Multipath]\label{thm::CRB::TOA::Coh::Comb}
Denoting the TOA estimate as $\hat{\tau}_{d,b}$, the variance of the TOA estimate between the target UE and anchor \(b\) can be lower-bounded as:
\begin{align}
    & {\rm var}(\hat{\tau}_{d,b}) \geq  \mathbb{I}_{e,\theta_{\hat{\tau}}^b}^{-1},\label{eq:cirbasedcrlb_toa}
\end{align}
where vector parameter $\theta_{\hat{\tau}}^b$ comprises $\tau_{d,b}$ and $\boldsymbol{h}_b$ to facilitate the joint estimation of TOA and channel state information (CSI) and $\mathbb{I}_{e,\theta_{\hat{\tau}}^b}$ is the effective FIM (EFIM) representing TOA information. EFIM \(\mathbb{I}_{e,\theta_{\hat{\tau}}^b}\) can be derived as
\begin{align}
    \mathbb{I}_{e,\theta_{\hat{\tau}}^b} & = 2\gamma_b{\boldsymbol{h}_b}^H \boldsymbol{F}_L^H {\boldsymbol{X}_b}^H \boldsymbol{D} \boldsymbol{\Xi}_b \boldsymbol{D} {\boldsymbol{X}_b} \boldsymbol{F}_L \boldsymbol{h}_b,\label{eq:coherent_fim}\\
    \boldsymbol{\Xi}_b & = \boldsymbol{I}_N-\boldsymbol{X}_b \boldsymbol{F}_L \big(\boldsymbol{F}_L^H {\boldsymbol{X}_b}^H \boldsymbol{X}_b \boldsymbol{F}_L\big)^{-1}\boldsymbol{F}_L^H {\boldsymbol{X}_b}^H,\label{eq:Xi_fim}
\end{align}
where \(\gamma_b = \frac{P_b}{\sigma^2}\) is the SNR of the received signal, $\boldsymbol{D}={\rm diag}\big(2\pi\nabla f(-\frac{N}{2}),\dots, 2\pi\nabla f(\frac{N}{2}-1)\big)$, and \(\boldsymbol{I}_N\) is the identity matrix of size \(N\times N\).
\end{theorem}
Due to space constraints, we cannot provide detailed proof in this paper. A proof sketch can be found in one of our previous conference papers~\cite{10632801_harish_dyspan}.

We now derive the variance of the TOA estimate for an AWGN channel, i.e., the LOS scenario, as a special case of Theorem~\ref{thm::CRB::TOA::Coh::Comb}. For the AWGN channel, the CIR \(\boldsymbol{h}_b\) in~\eqref{eq:coherent_fim} reduces to a single path. Assuming Quadrature Phase Shift Keying (QPSK) symbols for PRS, which aligns with 5G NR~\cite{3gpp::38211}, we have \(\boldsymbol{X}_b\boldsymbol{X}_b^H = \boldsymbol{X}_b^H\boldsymbol{X}_b = \boldsymbol{I}_N\). Similarly, the DFT matrix \(\boldsymbol{F}_L\) satisfies \(\boldsymbol{F}_L^H\boldsymbol{F}_L = \boldsymbol{I}_L\). Moreover, for large \(N\), we can approximate \(\boldsymbol{F}_L\boldsymbol{F}_L^H\) as \(\mathbf{0}_N\), a zero matrix of size \(N \times N\). Based on these results, we observe that \(\boldsymbol{\Xi}_b\) in~\eqref{eq:Xi_fim} approaches \(\boldsymbol{I}_N\). Using these simplifications, we present below the CRLB for TOA under AWGN as a corollary.

\begin{corollary}[CRLB of TOA for AWGN]\label{cor::CRB:TOA::AWGN}
The variance of the TOA estimate between the target UE and anchor \(b\) under the AWGN channel is lower-bounded as:
\begin{align}
    {\rm var}(\hat{\tau}_{d,b}) & \geq \mathbb{I}_{e,\theta_{\hat{\tau}}^b}^{-1},\label{eq:awgnbasedcrlb_toa}\\
    \mathbb{I}_{e,\theta_{\hat{\tau}}^b} & = 2\gamma_b\boldsymbol{D}^2 = 8\pi\gamma_b\nabla f^2\sum_{n=-\frac{N}{2}}^{\frac{N}{2}-1}n^2.\label{eq:fim_awgn}
\end{align}
\end{corollary}

Having obtained the CRLB for the variance of the TOA error and denoting the vector parameter for TDOA estimation as $\boldsymbol{\theta}_{\tilde{\tau}}$, the FIM for TDOA is expressed as
\begin{align}
    \mathbb{I}_{\theta_{\tilde{\tau}}} = \big(\mathbf{H}_{\hat{\tau}\rightarrow\tilde{\tau}}^T \mathbb{I}_{\theta_{\hat{\tau}}}^{-1} \mathbf{H}_{\hat{\tau}\rightarrow\tilde{\tau}}\big)^{-1},\label{eq:fim_tdoa_from_toa}
\end{align}
where $\mathbf{H}_{\hat{\tau}\rightarrow\tilde{\tau}}$ is a transformation matrix from TOA to TDOA.

Then, the FIM for position parameter $\boldsymbol{\theta}_p$ using TOA-based localization (denoted by $\mathbb{I}_{\hat{\boldsymbol{\theta}}_p}$) and TDOA-based localization (denoted by $\mathbb{I}_{\tilde{\theta}_p}$) can be obtained as follows
\begin{align}
    \mathbb{I}_{\hat{\boldsymbol{\theta}}_p} & = \mathbf{H}_{\hat{\tau}\rightarrow\boldsymbol{\theta}_p}^T \mathbb{I}_{\theta_{\hat{\tau}}} \mathbf{H}_{\hat{\tau}\rightarrow\boldsymbol{\theta}_p},\,
    \mathbb{I}_{\tilde{\theta}_p} = \mathbf{H}_{\tilde{\tau}\rightarrow\boldsymbol{\theta}_p}^T \mathbb{I}_{\theta_{\tilde{\tau}}} \mathbf{H}_{\tilde{\tau}\rightarrow\boldsymbol{\theta}_p}, \label{eq:fim_theta_toa_tdoa}
\end{align}
where $\mathbf{H}_{\hat{\tau}\rightarrow\boldsymbol{\theta}_p}$ and $\mathbf{H}_{\tilde{\tau}\rightarrow\boldsymbol{\theta}_p}$ are the transformation matrices from TOA and TDOA measurements to the position parameter, respectively. Denoting position estimate by \(\hat{\boldsymbol{\theta}}_p\) for TOA-based localization and \(\tilde{\boldsymbol{\theta}}_p\) for TDOA, the variance of the position estimate, also known as position error bound (PEB), is then given by
\begin{align}
    {\rm var}(\hat{\boldsymbol{\theta}}_p) & \geq \text{trace}(\mathbb{I}^{-1}_{\hat{\boldsymbol{\theta}}_p}),\,
    {\rm var}(\tilde{\boldsymbol{\theta}}_p) \geq \text{trace}(\mathbb{I}^{-1}_{\tilde{\boldsymbol{\theta}}_p}). \label{eq:crlb_peb_toa_tdoa}
\end{align}

\subsection{NLOS Bias Modeling}\label{sec:nlos_modeling}
To address the limitations of traditional NLOS bias modeling, which relies on Gamma-distribution-based methods, we propose a novel approach for modeling NLOS bias. As part of this, we first define TOA measurements under AWGN and multipath propagation scenarios; using these, we will provide a more realistic representation of NLOS bias.

Without loss of generality, the TOA measurement between an anchor and a target UE under the AWGN channel is expressed as 
\begin{align}
\hat{\tau}_d = \tau_d + v, \label{eq:toa_awgn}
\end{align} 
where \(\tau_d\) represents the true range between the anchor and the target UE, \(v\) is AWGN with zero mean and variance specified in~\eqref{eq:awgnbasedcrlb_toa}, and \(\hat{\tau}_d\) is the measured range. The variance of \(\hat{\tau}_d\), denoted as \(\sigma_n^2\), is equivalent to the variance of \(v\). Thus,  
\begin{align}
    \sigma_n^2 = {\rm var}(v) \geq \frac{1}{8\pi\gamma\nabla f^2\sum_{n=-\frac{N}{2}}^{\frac{N}{2}-1}n^2}, \label{eq:var_toa_awgn}
\end{align}  
where \(\gamma\) denotes the SNR of the considered link. In Monte Carlo simulations, TOA errors for AWGN channels can be modeled by adding AWGN noise with variance \(\sigma_n^2\) to the true range as follows
\begin{align}
\hat{\tau}_d = \tau_d + u, \quad u\sim\mathcal{CN}(0, \sigma_n^2).\label{eq:awgn_monte_carlo}
\end{align}

Similarly, TOA measurements between an anchor and a target UE under multipath propagation are defined as
\begin{align}
    \tilde{\tau}_d = \tau_d + b + v, \label{eq:toa_cir}
\end{align}  
where \(b\) is the NLOS bias induced by multipath propagation with non-zero mean and finite variance. The measured range, \(\tilde{\tau}_d\), incorporates the NLOS bias and AWGN, with variance specified in~\eqref{eq:cirbasedcrlb_toa}. The variance of \(\tilde{\tau}_d\), denoted as \(\sigma_c^2\), can be expressed as  
\begin{align}
    \sigma_c^2 = {\rm var}(b + v) = {\rm var}(b) + {\rm var}(v) = \sigma_b^2 + {\rm var}(\hat{\tau}_d),
\end{align}  
where \(\sigma_b^2\) is the variance of \(b\). Using this, we model NLOS bias offset \(\sigma_b\) as  
\begin{align}
\sigma_b = \sqrt{{\rm var}(\tilde{\tau}_d) - {\rm var}(\hat{\tau}_d)}.\label{eq:novel_nlos_bias}
\end{align}

Note that \(\sigma_b\) differs for each anchor-UE link. To model TOA errors for Monte Carlo simulations in the presence of multipath, NLOS bias offset \(\sigma_b\) and AWGN are added to the true range with the variance \(\sigma_c^2\) as follows
\begin{align}
\tilde{\tau}_d = \tau_d + \sigma_b + u, \quad u\sim\mathcal{CN}(0, \sigma_c^2).\label{eq:cir_monte_carlo}
\end{align}

\begin{figure}[t]
    \centering
    \includegraphics[width=0.90\linewidth]{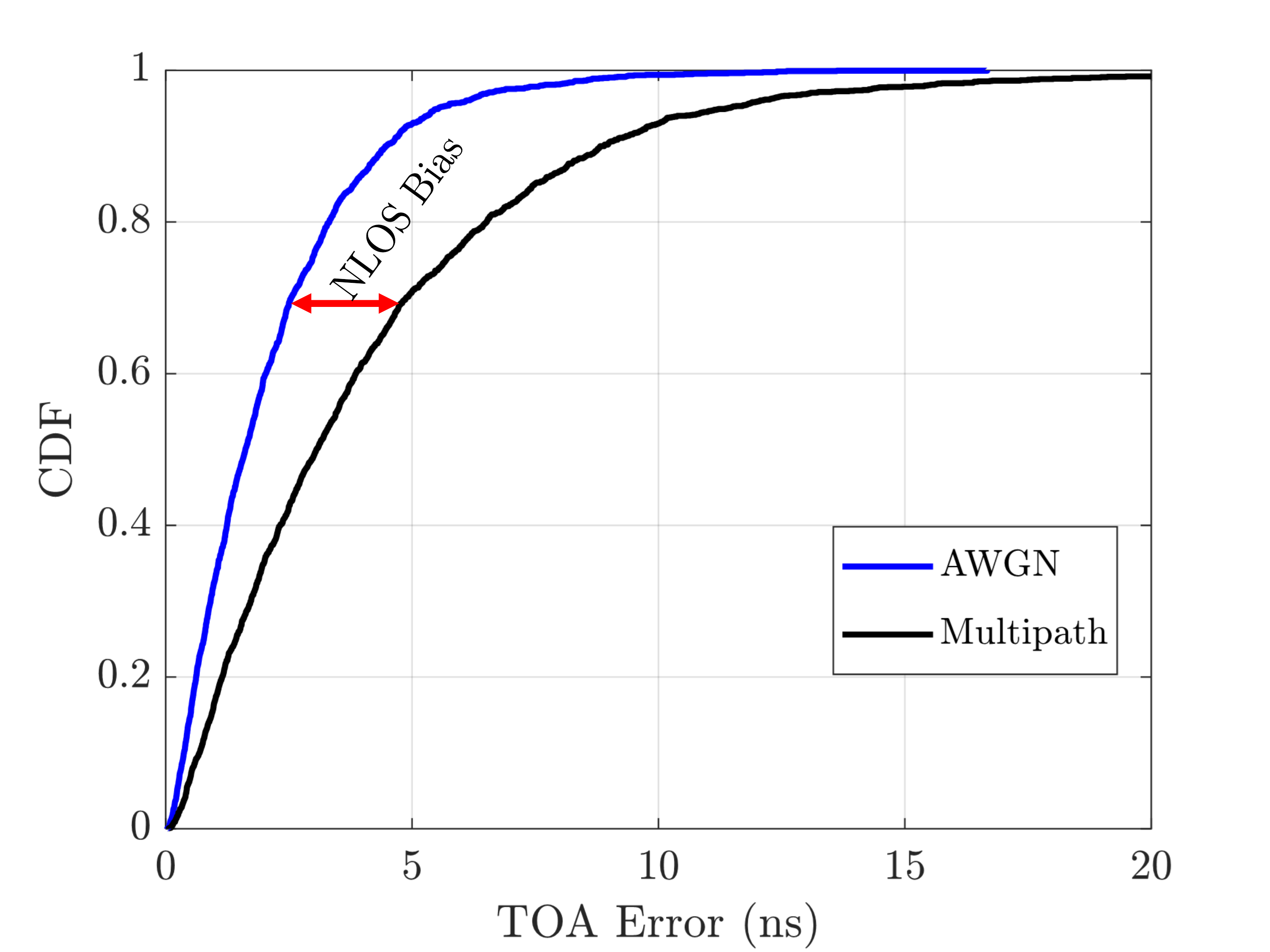}
    \captionsetup{font=small}
    \caption{{\em TOA error comparison}: Comparing TOA error observed in AWGN and multipath.}
    \label{fig:toa_err_awgn_cir}
\vspace{-15pt} \end{figure}

By explicitly linking the NLOS bias offset to the SNR, multipath effects, and physical layer characteristics, our proposed approach provides a more realistic and comprehensive representation of the NLOS bias. The proposed approach bridges the gap between theoretical analyses and practical channel conditions, ensuring improved accuracy in modeling and simulation. To further illustrate NLOS bias modeling, we assess the TOA error performance with an anchor-UE drop presented in Fig.~\ref{fig:anchor_ue_drop1}. Fig.~\ref{fig:toa_err_awgn_cir} shows a comparison of the TOA error observed under AWGN (described in~\eqref{eq:awgnbasedcrlb_toa}) and the multipath propagation (as given in~\eqref{eq:cirbasedcrlb_toa}). As anticipated, the TOA error observed under multipath is higher due to NLOS bias offsets. The difference between TOA errors observed in Fig.~\ref{fig:toa_err_awgn_cir} relates to the NLOS bias offset derived in~\eqref{eq:novel_nlos_bias}. After establishing the analytical framework and NLOS modeling, we will proceed to assess positioning performance in the next section.

\section{Results and Discussion}\label{sec:results_discussion}

\begin{figure}[t]
    \centering
    \includegraphics[width=0.90\linewidth]{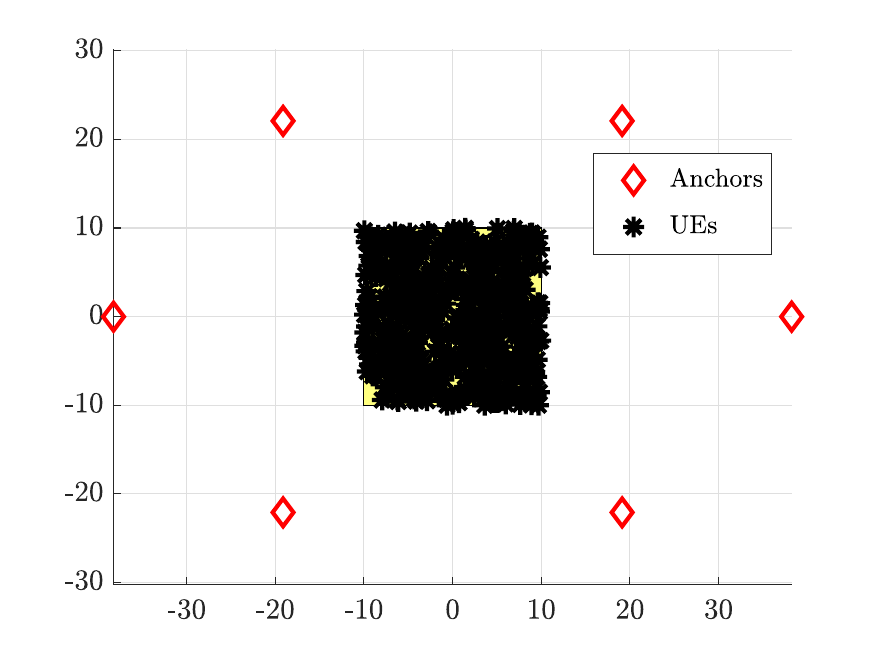}
    \captionsetup{font=small}
    \caption{Anchor - UE drop: Good Geometry.}
    \label{fig:anchor_ue_drop1}
\vspace{-18pt} \end{figure}

In this section, we evaluate the positioning performance of the proposed TS-WPM and benchmark it against CRLB. We begin by outlining the evaluation framework. Our evaluation considers the anchor-UE deployment scenario illustrated in Fig.~\ref{fig:anchor_ue_drop1}, where UEs are uniformly distributed within a predefined rectangular area, while anchors are strategically positioned around this area to optimize coverage and geometric diversity. In this configuration, each UE attempts to measure TOA from all visible anchors and subsequently computes TDOA measurements. TDOA measurements are computed using the anchor with the strongest SNR as the reference anchor. For this study, we assume that all UEs are located indoors, leading to predominantly NLOS conditions between the anchors and UEs. This aligns with 3GPP evaluation scenarios~\cite{3gpp::38901,3gpp::36814} and avoids biases that could arise from scenarios favoring higher accuracy. The dual-stripe path loss model~\cite{3gpp::36814} is employed to account for distance-dependent path loss and penetration losses. To capture realistic channel characteristics, shadow fading effects are also considered~\cite{3gpp::36814}. To incorporate small-scale fading, we utilize clustered delay line (CDL) channel models, specifically the CDL-A channel model, which is well-suited for NLOS scenarios~\cite{3gpp::38901}. The system operates at a carrier frequency of 3.5 GHz with a 5 MHz bandwidth. We assume a maximum transmit (Tx) power of 23 dBm. This study demonstrates the effectiveness of our proposed TS-WPM by comparing it with state-of-the-art approaches, including IPPM, NLS, and WNLS, with all results benchmarked against the CRLB. \chb{To provide a comprehensive performance benchmark, we also evaluate TS-WPM under an AWGN channel. This aligns with the closed-form CRLB expression derived in Corollary~\ref{cor::CRB:TOA::AWGN} and facilitates quantifying the performance degradation in multipath-rich environments by comparison with the generalized CRLB derived in Theorem~\ref{thm::CRB::TOA::Coh::Comb}.} Additionally, we present a computational complexity analysis to compare the efficiency of TS-WPM with these state-of-the-art approaches.  

\chb{The said baselines were chosen for the following reasons: 1) The proposed TS-WPM is an iterative approach that updates location estimates in a manner similar to IPPM, NLS, and WNLS methods, making these comparisons particularly relevant. 2) Optimization-based techniques such as SDP~\cite{9376594,6268731,8241388} and MDS~\cite{10.1145/1138127.1138129,5342507}  are typically designed for TOA-only settings, offer moderate accuracy, are computationally complex, and do not offer optimal performance even under AWGN, unlike WNLS. 3) Machine learning-based methods~\cite{Bhatti2018MachineLB,tian2023highprecisionmachinelearningbasedindoor} require large datasets,  extensive offline training, and lack analytical bounds. Moreover, environment-specific training/validation limits their generalizability, e.g., low SNR or high GDOP scenarios. 4) WNLS remains one of the most sophisticated and widely adopted estimators for TDOA-based localization, offering near-optimal performance under favorable conditions~\cite{10.1007/s11277-007-9375-z,10.5555/555289}. Furthermore, several recent works~\cite{ZHUO2024,8947090_liu_weight_ippm,8979252_liu_covar_ippm} have explored low-complexity localization strategies including greedy search, weighted PPM, and covariance-aware refinement. While promising, these methods are limited to TOA-only settings and do not support TDOA, and evaluation under realistic conditions, such as multipath and radio propagation aspects. Thus, by considering WNLS as the strongest conventional benchmark and demonstrating the superiority and near-CRLB performance of TS-WPM across various conditions (e.g., high GDOP, low SNR, multipath), our evaluation implicitly establishes the robustness and practical relevance of TS-WPM, even when compared to the latest localization techniques. Note that the IPPM baseline included in this study is the reformulated version we proposed in our earlier work~\cite{10632801_harish_dyspan}, where we extended traditional IPPM to support TDOA-based localization. As shown in~\cite{10632801_harish_dyspan}, while this extended IPPM offers lower complexity, its performance is comparable to or matches that of NLS at best, a finding further reinforced through our results in this section.} 

\chb{Moreover, in all experiments, we set the convergence threshold to \(\epsilon = 10^{-7}\), and require the residual error to remain below this threshold for \(l = 10\) consecutive iterations before terminating the iterative loop and declaring convergence. While \(\epsilon = 10^{-7}\) enables high precision convergence, the choice of \(l = 10\) prevents premature termination due to transient noise, thereby ensuring convergence stability. Additionally, we put a hard limit on the total number of iterations at 100 to ensure computational feasibility and bounded complexity, especially under high-GDOP or low-SNR conditions. These values were determined through empirical validation across diverse deployment scenarios, providing a robust trade-off between accuracy and efficiency.}
\begin{figure*}[t]
     \centering
     \begin{subfigure}[b]{0.44\textwidth}
         \centering
         \includegraphics[width=\textwidth]{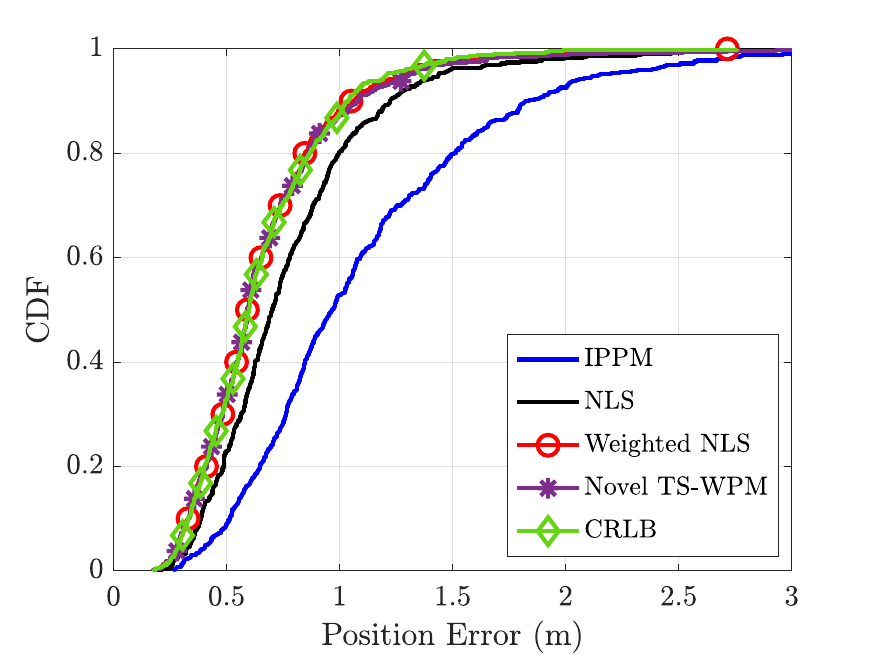}
         \captionsetup{font=small}
         \caption{Position Error: AWGN Channel.}
         \label{fig:cdf_poserr_23dbm_low_gdop_awgn}
     \end{subfigure}
     \hfill
     \begin{subfigure}[b]{0.44\textwidth}
         \centering
        \includegraphics[width=\textwidth]{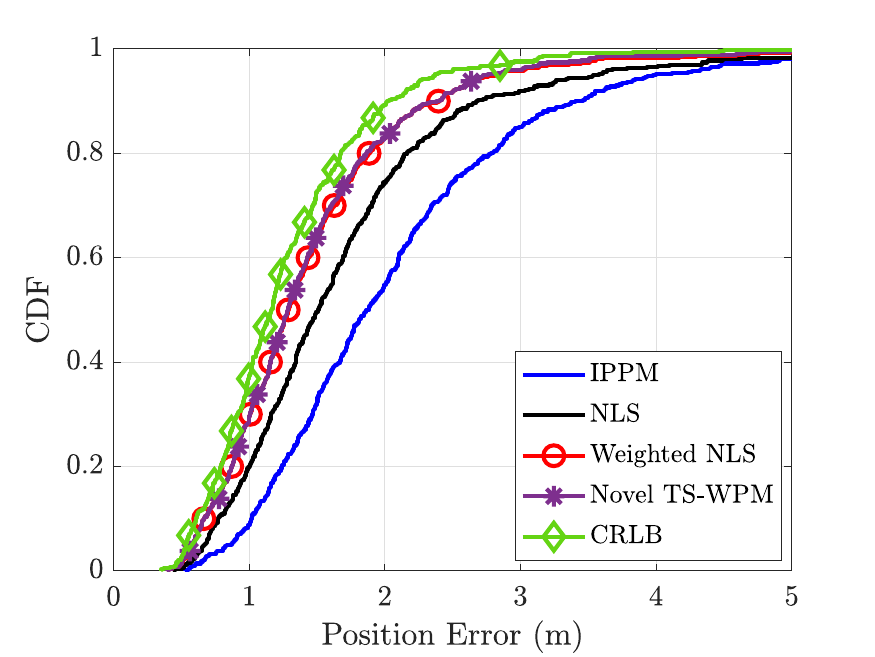}
        \captionsetup{font=small}
        \caption{Position Error: Multipath Channel.}
        \label{fig:cdf_poserr_23dbm_low_gdop_cir}
     \end{subfigure}
     \captionsetup{font=small}
     \caption{CDF of position error evaluated at 3.5 GHz with 5 MHz bandwidth in AWGN and multipath: Good geometry and high SNR regime}
     \label{fig:poserr_23dbm_low_gdop}
\vspace{-18pt} \end{figure*}

\subsection{Positioning Performance}
Fig.~\ref{fig:cdf_poserr_23dbm_low_gdop_awgn} and Fig.~\ref{fig:cdf_poserr_23dbm_low_gdop_cir} present the positioning performance for the anchor-UE deployment depicted in Fig.~\ref{fig:anchor_ue_drop1}, evaluated under both AWGN and multipath propagation scenarios. The results are obtained through extensive Monte Carlo simulations, where TOA errors are modeled in accordance with~\eqref{eq:awgn_monte_carlo} and~\eqref{eq:cir_monte_carlo}. As illustrated, in the AWGN scenario, our novel approach exhibits positioning accuracy superior to IPPM and NLS and matches the performance of WNLS. This observation is significant as WNLS is well-known for achieving CRLB in AWGN scenarios~\cite{buehrer2019handbook}, a standard of optimality that is also met by our two-stage method. A similar trend is observed when extending the evaluation to multipath propagation conditions. Notably, even under multipath conditions, where NLOS effects and biases are prevalent, the performance of the proposed method remains remarkably close to the CRLB. This result also highlights that, under favorable geometry and high SNR conditions, both our proposed approach and WNLS are asymptotically optimal.
\begin{figure}
    \centering
    \includegraphics[width=0.90\linewidth]{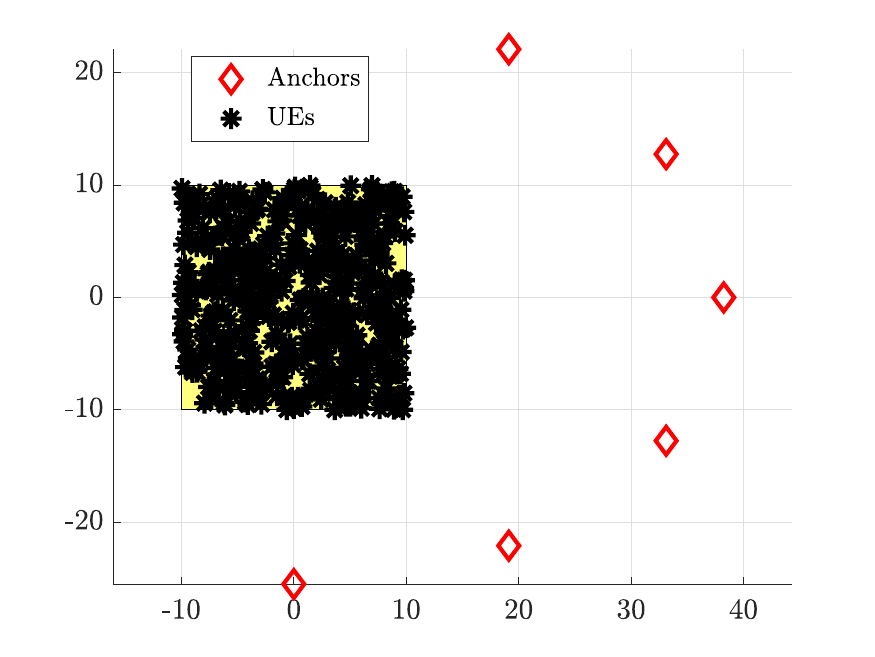}
    \captionsetup{font=small}
    \caption{Anchor - UE drop: Non-ideal Geometry.}
    \label{fig:anchor_ue_drop2}
\vspace{-18pt} \end{figure}
\begin{figure}[t]
     \centering
     \begin{subfigure}[b]{0.90\linewidth}
         \centering
         \includegraphics[width=\linewidth]{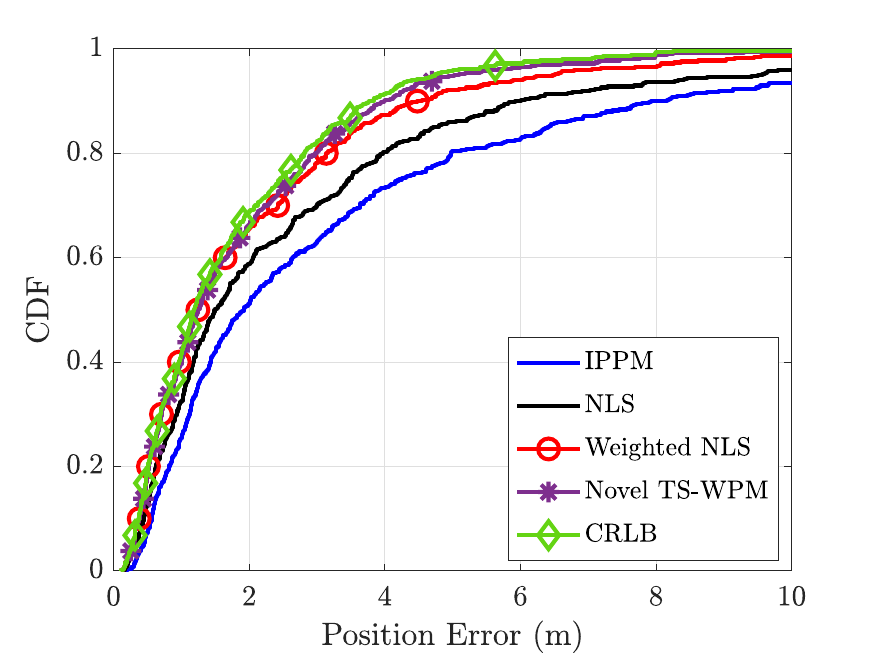}
         \captionsetup{font=small}
         \caption{Position Error: AWGN Channel.}
         \label{fig:cdf_poserr_23dbm_high_gdop_awgn}
     \end{subfigure}
     \hfill
     \begin{subfigure}[b]{0.90\linewidth}
         \centering
        \includegraphics[width=\linewidth]{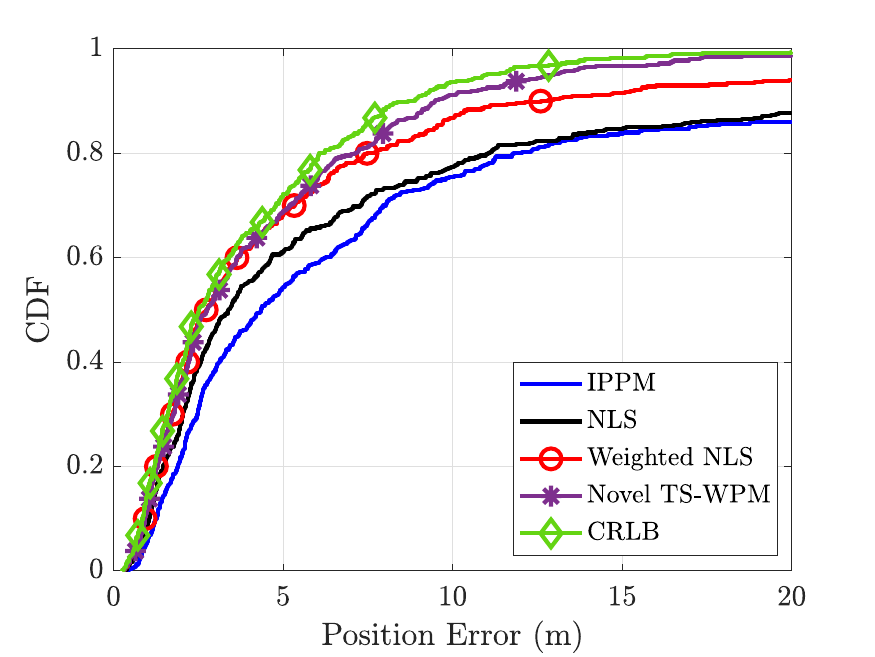}
        \captionsetup{font=small}
        \caption{Position Error: Multipath Channel.}
        \label{fig:cdf_poserr_23dbm_high_gdop_cir}
     \end{subfigure}
     \captionsetup{font=small}
     \caption{CDF of position error evaluated at 3.5 GHz with 5 MHz bandwidth in AWGN and multipath: Non-ideal geometry}
     \label{fig:poserr_23dbm_high_gdop}
\vspace{-18pt} \end{figure}

To demonstrate the superior performance of TS-WPM over WNLS, we evaluate positioning performance under non-ideal geometry conditions. For this evaluation, we use anchor-UE deployment shown in Fig.~\ref{fig:anchor_ue_drop2}. In this configuration, while UE distribution remains consistent with the earlier evaluation, anchor deployment is configured such that the anchor geometry is worse compared to the earlier evaluation, thereby increasing the GDOP. All other simulation assumptions remain consistent with the previous assessment. Fig.~\ref{fig:cdf_poserr_23dbm_high_gdop_awgn} and Fig.~\ref{fig:cdf_poserr_23dbm_high_gdop_cir} present the positioning performance under both AWGN and multipath propagation for this non-ideal geometry scenario. As observed, the proposed TS-WPM continues to outperform IPPM and NLS. More importantly, we observe that TS-WPM surpasses WNLS across both AWGN and multipath conditions. The performance gap is more pronounced in the multipath scenario, where NLOS bias offsets, coupled with high GDOP, further degrade the accuracy of WNLS. The improved performance of TS-WPM, especially under these non-ideal conditions, can be attributed to the two-stage refinement process combined with the modified normalization factor, as detailed in Section~\ref{sec:analytical_study_tswpm}.

\begin{figure}[t]
     \centering
    \includegraphics[width=0.90\linewidth]{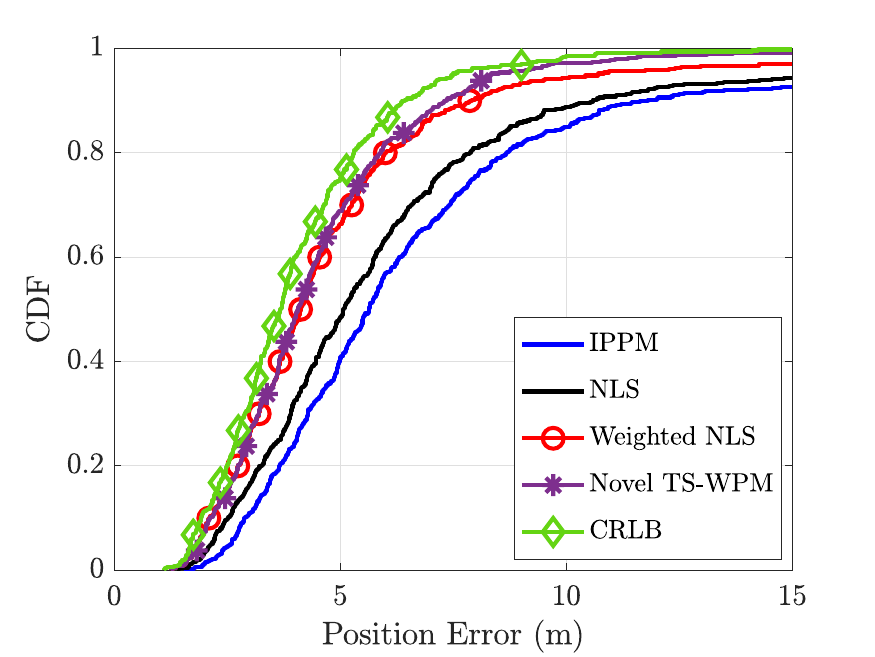}
     \captionsetup{font=small}
     \caption{CDF of position error evaluated at 3.5 GHz with 5 MHz bandwidth in multipath: Good geometry and low SNR regime}
    \label{fig:cdf_poserr_13dbm_low_gdop_cir}     
\vspace{-15pt} \end{figure}

To further demonstrate the superior performance of TS-WPM, we investigated another challenging scenario characterized by low SNR conditions. For this evaluation, we reuse the anchor-UE deployment presented in Fig.~\ref{fig:anchor_ue_drop1}, but instead of assuming a Tx power of 23 dBm, we reduce it to 13 dBm to induce low SNR conditions. All other simulation assumptions remain consistent with the previous assessment. Our evaluations indicated that, under AWGN conditions, TS-WPM performance closely matches that of WNLS; hence, we focus our discussion on the positioning performance observed under multipath propagation for low SNR scenarios. Fig.~\ref{fig:cdf_poserr_13dbm_low_gdop_cir} illustrates the positioning performance for low SNR under multipath propagation. It is clearly evident that TS-WPM demonstrates superior performance compared to WNLS, in addition to IPPM and NLS. This superior performance is attributed to the amplified errors in TOA measurements at low SNRs, which are further exacerbated due to multipath propagation, where the NLOS bias offset depends on SNR and the underlying propagation conditions. To better understand this phenomenon, consider TOA estimation under multipath propagation for both high and low SNR scenarios. In high SNR conditions, the TOA estimation process is generally able to lock onto the first path or a path close to the first one, as these paths are distinct and detectable. \chb{Conversely, under low SNR conditions, the first-arriving path may become insignificant or undetectable--not because only the first path is degraded, but because in multipath-rich environments (e.g., indoor settings), the first path is often weaker than subsequent reflected paths. Due to rich scattering, it may carry less energy and is more likely to fall below the detection threshold used in practical TOA estimation algorithms, such as correlation-based methods. In contrast, later-arriving stronger multipath components may still exceed the threshold and be detected, thereby biasing the TOA estimate toward delayed paths.}

Although our evaluation framework does not explicitly estimate TOA, the TOA error modeling in~\eqref{eq:awgn_monte_carlo} and~\eqref{eq:cir_monte_carlo} captures these effects by accounting for the prevailing SNR and propagation characteristics. Consequently, when TOA errors are significant--primarily due to a pronounced NLOS bias offset--the performance of WNLS deteriorates. In contrast, as described in Section~\ref{sec:novel_ts_wpm_tdoa}, the proposed approach incorporates a two-stage refinement process, which mitigates the impact of the NLOS bias offset by accurately estimating the range measurements associated with the reference anchor range and iteratively updating the UE position estimate and vice versa. As a result, the proposed approach ensures enhanced positioning even under low SNR conditions. Remarkably, even in high GDOP or low SNR scenarios, the proposed method achieves performance levels very close to the CRLB. \chb{Notably, the absolute magnitude of positioning errors is influenced by the system parameters used for evaluation, such as carrier frequency and bandwidth. While our evaluation adopts a representative 5G NR-compliant configuration (3.5 GHz carrier frequency with 5 MHz bandwidth) for demonstration purposes, the proposed TS-WPM is agnostic to these parameters and would yield significantly improved accuracy under wider bandwidths (e.g., 20-100 MHz) and lower carrier frequencies, as supported in 5G NR and beyond.}

\begin{figure}[t]
     \centering
    \includegraphics[width=0.90\linewidth]{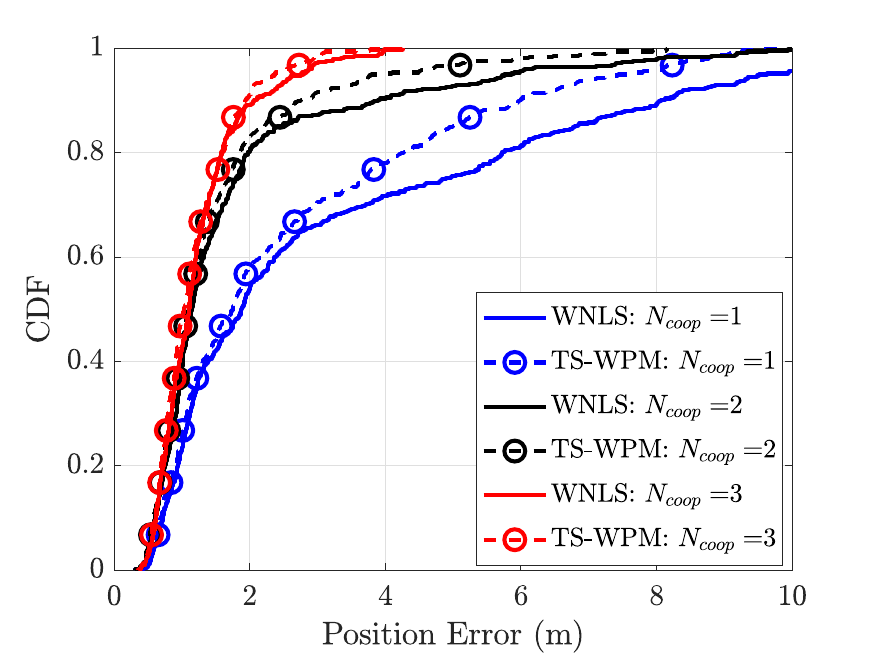}
     \captionsetup{font=small}
     \caption{{\em Performance with co-operative localization:} CDF of position error evaluated at 3.5 GHz with 5 MHz bandwidth in multipath}
     \label{fig:cdf_poserr_coop_cir}
\vspace{-18pt} \end{figure}

In our final evaluation, we assess the positioning performance of cooperative localization as outlined in Algorithm~\ref{alg:novel_ts_wpm_coop}. For this assessment, we reuse the anchor-UE deployment depicted in Fig.~\ref{fig:anchor_ue_drop1}, considering multipath propagation. However, unlike previous scenarios, we assume that each target UE has visibility to only two anchors, making collaborative measurements essential for accurate localization. The number of cooperative nodes, denoted by \(N_{coop}\), is set to \([1,\,2,\,3]\), with nodes randomly selected from the UE drop. Fig.~\ref{fig:cdf_poserr_coop_cir} illustrates the positioning performance of TS-WPM across different \(N_{coop}\). The results clearly demonstrate the effectiveness of cooperative localization in enabling accurate positioning when an insufficient number of anchors are visible, with increasing the number of cooperative nodes leading to further accuracy improvements. Additionally, we compare the results with WNLS across different \(N_{coop}\). It is crucial to note that WNLS, in its original form, fails to provide meaningful positioning estimates as the matrix in the normalization (see~\eqref{eq:weighted_nls_final_est}) becomes ill-conditioned (non-invertible) due to highly unfavorable GDOP resulting from randomly selected collaborative nodes combined with limited anchor availability. This aligns with the insights established in Proposition~\ref{prop::robustness::tswpm}, which show that WNLS results in high MSE in high GDOP scenarios. Addressing this, we employed a regularization factor to stabilize the matrix inversion process, a concept that is inherently incorporated in TS-WPM through a more systematic approach, as discussed in Section~\ref{sec:alt_math_intuition}. Our evaluation confirms that TS-WPM consistently outperforms WNLS due to its sophisticated normalization coupled with the two-stage approach, reinforcing its robustness under high GDOP conditions. Furthermore, we observed that cooperative localization with IPPM, under this setup, resulted in prohibitively high positioning errors, which is why we opted not to include a comparison with IPPM in this scenario. These findings underscore the effectiveness of TS-WPM in tackling challenging localization conditions where conventional approaches struggle to maintain accuracy.

\subsection{Computational Complexity}
Having established the superior positioning performance of TS-WPM, we now analyze the computational complexity of the proposed TS-WPM and compare it with state-of-the-art algorithms, including NLS, WNLS, and IPPM. As outlined in Algorithm~\ref{alg:novel_ts_wpm_tdoa}, each iteration of TS-WPM involves updating the UE location estimate, computing residuals, and refining the reference anchor range estimate across all \(B\) anchors (see steps~\ref{step:novel_ts_wpm_tdoa_loc_update}, \ref{step:novel_ts_wpm_tdoa_res_err}, and \ref{step:novel_ts_wpm_toa_range_update}). Each update step comprises basic vector norms and arithmetic operations, resulting in a per-iteration complexity of \(O(B\cdot D)\). Consequently, the overall complexity of TS-WPM is \(O(N\cdot B\cdot D)\), where \(N\) represents the number of iterations required for convergence and \(D\) is the dimension of UE location coordinates (2 in our case). Similarly, the computational complexity of IPPM is also \(O(N\cdot B\cdot D)\)~\cite{5677549_tao_coop_ippm,5683693_toa_coop_mod_ippm}, which is identical to that of TS-WPM. In contrast, the computational complexity of NLS and WNLS is significantly higher due to additional matrix inversion and multiplication operations. Specifically, both methods exhibit a complexity of \(O(N \cdot B \cdot D^4)\)~\cite{buehrer2019handbook}, where \(D\) denotes the number of spatial dimensions of the UE coordinates. NLS and WNLS typically require fewer iterations than IPPM and TS-WPM, with TS-WPM necessitating additional iterations due to its two-stage refinement process. Our evaluations indicate that, on average, NLS and WNLS converge in 16-17 iterations, while IPPM and TS-WPM require more than 20 iterations. Despite its higher iteration count, TS-WPM remains computationally efficient compared to NLS and WNLS, as it avoids costly matrix inversions and minimizes the number of matrix multiplications.

\chb{{\em Practical Considerations and Generalizability.} While our evaluation framework assumes the availability of TOA/TDOA measurements, we acknowledge that in real-world deployments, TOA measurements are often subject to imperfections such as multipath-induced bias (NLOS bias), thermal noise, anchor synchronization errors, and residual UE clock offsets. To address these, our framework explicitly incorporates the effects of thermal noise and multipath through 3GPP-compliant CRLB formulations and the associated TOA error models, including a novel NLOS bias model that accounts for radio propagation characteristics and SNR, as described in Section~\ref{sec:analytical_eval_framework}. Moreover, the TDOA formulation inherently eliminates UE clock offset errors, assuming that the TOA measurements are acquired within a narrow measurement window (e.g., a few ms), during which the UE clock remains stable. While the current model assumes perfect synchronization among anchors, this is a valid assumption in many practical scenarios, such as 5G gNBs, Global Navigation Satellite System (GNSS)-synchronized UAV-mounted gNBs, or Low Earth Orbit (LEO) satellites equipped with GNSS receivers. Importantly, our framework is flexible enough to allow seamless integration of practical TOA estimation methods (e.g., correlation-based, Multiple Signal Classification (MUSIC), and Estimation of Signal Parameters via Rotational Invariant Techniques (ESPRIT))--as well as clock offset modeling and compensation mechanisms.}

\section{Concluding Remarks}\label{sec:conclusion}
In this paper, we proposed a novel low-complexity TS-WPM for TDOA-based localization and extended it to support cooperative localization. The proposed method employs a two-stage refinement process that iteratively updates both the range estimate associated with the reference anchor and the UE location estimate, thereby significantly enhancing positioning accuracy while maintaining computational efficiency. This refinement process proves particularly effective in challenging scenarios such as high GDOP and low SNR. Furthermore, we provided an analytical derivation establishing the conditions under which TS-WPM is asymptotically optimal and demonstrating its performance advantages over WNLS. To rigorously evaluate the proposed method, we developed a comprehensive 3GPP-compliant analytical framework that integrates physical layer and fading characteristics, using the CRLB as a principal tool. Additionally, we proposed a novel NLOS bias modeling approach that explicitly accounts for underlying propagation conditions and SNR, offering a more accurate representation of the bias induced by multipath. Through extensive Monte Carlo simulations, we demonstrated that the proposed TS-WPM consistently outperforms state-of-the-art techniques such as IPPM and NLS. Our results illustrated that, under non-ideal conditions, TS-WPM surpasses WNLS, achieving performance close to the CRLB even in scenarios with high GDOP, low SNR, and multipath propagation. Furthermore, we highlighted the effectiveness of the proposed approach when incorporating cooperative localization, enabling accurate positioning even when an insufficient number of anchors (such as 2) are visible. Finally, we performed a computational complexity analysis comparing TS-WPM with IPPM, NLS, and WNLS, demonstrating that the proposed approach exhibits complexity identical to IPPM and lower complexity than NLS and WNLS, as it avoids complex operations like matrix inversion and minimizes matrix multiplications. Overall, the proposed TS-WPM presents a balanced trade-off between positioning accuracy and computational complexity, making it a promising candidate for real-time localization applications across diverse environments.

\appendix
\begin{chbgrp}
\subsection{Key Definitions and Evaluation Metrics}\label{app::key_definitions}
{\em Signal-to-Noise Ratio (SNR).} Based on the received signal model in (29), the SNR between anchor \(b\) and the target UE, denoted by \(\gamma_b\), is defined as
\begin{align}
    \gamma_b = \frac{P_b}{\sigma^2 + NF_l},
    \label{eq:snr_def_anchor_b}
\end{align}
where \(P_b\) represents the received signal power, \(\sigma^2\) is variance of AWGN, and \(NF_l\) denotes the noise figure in linear scale. To accurately compute \(P_b\) and, consequently, \(\gamma_b\), we adopt the dual-stripe channel model recommended in~\cite{3gpp::36814}. This model partitions the environment into line-of-sight (LOS) and NLOS regions, referred to as ``stripes". The stripe selection depends on the scenario--LOS stripes are used for outdoor UEs, while NLOS stripes are used for indoor UEs. Since our study assumes all UEs are located indoors, the NLOS stripe is employed to model large-scale fading.

The large-scale fading model accounts for distance-dependent path loss, shadowing, and penetration losses. The total path loss (in dB) is calculated as
\begin{align}
    PL & = \max(15.3 + 37.6\log_{10}R,\, 38.46 + 20\log_{10}R)  \notag\\
       &\quad 0.7d_{2D,\text{indoor}} + 18.3\,n^{\left(\frac{n+2}{n+1} - 0.46\right)} + L_{iw} + L_{ow},
    \label{eq:pathloss}
\end{align}
where \(R\) is the 3D distance between the anchor and the UE, \(d_{2D,\text{indoor}}\) is the horizontal indoor distance in meters, and \(n\) is the number of floors penetrated, capturing the impact of buildings with multiple floors. \(L_{iw}\) denotes the loss from internal walls (typically 5 dB per wall), and \(L_{ow}\) accounts for penetration through external walls (typically 20 dB per wall).

Given the path loss, the SNR in dB is expressed as
\begin{align}
    \gamma = P_t - PL - SF - N_0 - NF,
    \label{eq:snr_formula}
\end{align}
where \(P_t\) is the transmit power in dBm (assumed to be 23 dBm, except for the low SNR case where it is reduced to 13 dBm), \(PL\) is the path loss in dB as defined above, \(SF\) is the shadow fading loss in dB, \(NF\) is the noise figure in dB, and \(N_0\) is the thermal noise power in dBm computed as
\begin{align}
    N_0 = -174 + 10\log_{10}(B),
\end{align}
assuming a room temperature of \(T = 290\) K and a system bandwidth \(B\) in Hz. 
The SNR computed using~\eqref{eq:snr_formula} is used throughout our evaluation to quantify signal quality for each anchor-UE link.

{\em Geometric Dilution of Precision (GDOP).} GDOP is a fundamental metric that quantifies the impact of anchor geometry on the accuracy of location estimation. Even when range measurements are accurate, unfavorable anchor distribution can amplify positioning errors due to high GDOP. Conversely, well-distributed anchors result in low GDOP and improved localization performance. To formalize this, consider a network of anchors where \(\mathbf{r}\) denotes the true range measurements, \(\boldsymbol{\theta}\) is the position parameter comprising target UE coordinates, and \(\mathbf{H}\) is the corresponding Jacobian matrix relating the measurements to the UE location. The measurement model can be expressed as
\begin{align}
    \mathbf{r} = \mathbf{H}\boldsymbol{\theta}.
\end{align}

The GDOP is then defined by
\begin{align}
    {\rm GDOP} = \sqrt{{\rm trace}\left(\left(\mathbf{H}^T\mathbf{H}\right)^{-1}\right)},
\end{align}
where \((\cdot)^T\) denotes matrix transpose. In GPS settings~\cite{Edkgps2005}, the matrix \((\mathbf{H}^T\mathbf{H})^{-1}\) takes the form
\begin{align}
    \left(\mathbf{H}^T\mathbf{H}\right)^{-1} = \begin{bmatrix}
  D_{11} & D_{12} & D_{13} & D_{14}\\
  D_{21} & D_{22} & D_{23} & D_{24}\\
  D_{31} & D_{32} & D_{33} & D_{34}\\
  D_{41} & D_{42} & D_{43} & D_{44}
  \end{bmatrix}.
\end{align}
from which various DOP metrics, including position DOP (PDOP), horizontal DOP (HDOP), vertical DOP (VDOP), and time DOP (TDOP) are derived.
\begin{align}
    {\rm GDOP} & = \sqrt{D_{11}+D_{22}+D_{33}+D_{44}},\notag\\
    {\rm PDOP} & = \sqrt{D_{11}+D_{22}+D_{33}},\notag\\
    {\rm HDOP} & = \sqrt{D_{11}+D_{22}},{\rm VDOP} = \sqrt{D_{33}},\,{\rm TDOP} = \sqrt{D_{44}}.\notag
\end{align}
Here, PDOP quantifies the impact of anchor geometry on 3D positioning accuracy, HDOP on horizontal positioning accuracy, VDOP on vertical accuracy, and TDOP on the estimation of UE clock offset. GDOP reflects the overall geometric impact on all these components combined. For simplified scenarios with equal SNRs across anchors, the positioning error \(P_{\text{err}}\) can be approximated as a function of PDOP and range error \(r_{\text{err}}\) as
\begin{align}
    P_{\text{err}} = {\rm PDOP} \times r_{\text{err}}.
\end{align}

In this article, we focus on 2D localization assuming no UE clock offset errors (or assuming TDOA inherently cancels out UE clock offset errors); hence, HDOP is of primary interest and is treated equivalently to GDOP and PDOP in our evaluation. To illustrate the impact of anchor geometry, we compute the GDOP for the anchor-UE deployments used in our experiments. Fig.~\ref{fig:gdop_good_vs_non_ideal_geometry} presents the CDF of GDOP under two distinct geometric settings: a favorable anchor deployment (see Fig.~\ref{fig:anchor_ue_drop1}) and a non-ideal deployment (see Fig.~\ref{fig:anchor_ue_drop2}). As evident from the results, good geometry results in significantly lower GDOP values compared to non-ideal geometry, which is consistent with the positioning results presented earlier--compare the low-GDOP results in Fig.~\ref{fig:poserr_23dbm_low_gdop} with the degraded accuracy under high-GDOP conditions shown in Fig.~\ref{fig:poserr_23dbm_high_gdop}.
\begin{figure}[t]
    \centering
    \includegraphics[width=0.90\linewidth]{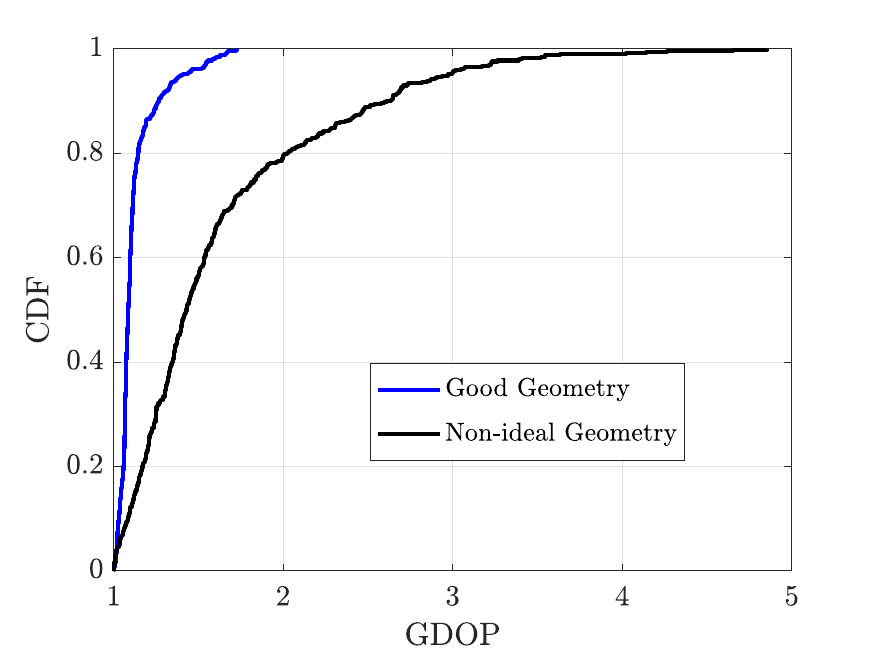}
    \captionsetup{font=small}
    \caption{\chb{{\em GDOP comparison}: Good geometry vs. non-ideal geometry.}}
    \label{fig:gdop_good_vs_non_ideal_geometry}
    \vspace{-15pt}
\end{figure}

To contextualize the GDOP values observed in our experiments, Table~\ref{tab:gdop_rating} provides the qualitative interpretation of DOP values, as originally presented in~\cite{Langley1999DOP}. A DOP value near 1 corresponds to ideal anchor geometry--such as when anchors are well spread--while values exceeding 10 typically reflect highly clustered or poorly distributed anchors, resulting in degraded localization accuracy. In our experiments, the GDOP values under favorable geometry consistently fall below 2, corresponding to the ``Excellent" rating. For the non-ideal geometry, GDOP values span both ``Excellent" and ``Good" ranges. We intentionally avoided scenarios with GDOP in the ``Fair" or ``Poor" ranges, as such configurations would severely degrade the performance of benchmark methods such as WNLS, making the comparison unfair and less meaningful. Instead, by selecting challenging yet realistic configurations (i.e., ``Good" vs. ``Excellent" geometry), we demonstrate that the proposed TS-WPM achieves superior performance even under moderate and not so unfavorable conditions.
\begin{table}[htbp]
\centering
\caption{\chb{GDOP Rating Classification}}
\begin{tabular}{|c|l|}
\hline
\textbf{GDOP Value} & \textbf{Rating Description} \\
\hline
1--2        & Excellent \\
2--5        & Good \\
5--10       & Fair \\
\(>\)10         & Poor \\
\hline
\end{tabular}
\label{tab:gdop_rating}
\end{table}
\end{chbgrp}

\subsection{Proof of Lemma~\ref{lemma::cov::mse::tswpm}}\label{app::cov::mse::tswpm}
By utilizing the expression for the final estimate of TS-WPM in~\eqref{eq:nwpm_final_est} and replacing \(\mathbf{H}_2(\boldsymbol{\theta})\) by \(\mathbf{H}_2\) for notational simplicity, we compute the covariance of the estimation error \( (\hat{\boldsymbol{\theta}}_{\mathrm{tswpm}}-\boldsymbol{\theta}) \) as follows
\begin{align}
    {\rm cov}(\hat{\boldsymbol{\theta}}_{\mathrm{tswpm}}-\boldsymbol{\theta}) = \mathbf{H}_2^T\mathbf{W}_2{\rm cov}(\tilde{\mathbf{r}} - \tilde{\mathbf{r}}'_e(\boldsymbol{\theta},r_1))\mathbf{W}_2\mathbf{H}_2.
    \label{eq:cov_nwpm_der1}
\end{align}

By substituting \( {\rm cov}(\tilde{\mathbf{r}} - \tilde{\mathbf{r}}'_e(\boldsymbol{\theta},r_1)) = \mathbf{C}_2 \) from~\eqref{eq:cov_tdoa_scen2} into~\eqref{eq:cov_nwpm_der1}, we obtain
\begin{align}
    {\rm cov}(\hat{\boldsymbol{\theta}}_{\mathrm{tswpm}}-\boldsymbol{\theta}) =\mathbf{H}_2^T\mathbf{W}_2\mathbf{C}_2\mathbf{W}_2\mathbf{H}_2.
    \label{eq:cov_nwpm_der2}
\end{align}

Next, by substituting \(\mathbf{W}_2 = \frac{\mathbf{C}_2^{-1}}{{\rm trace}(\mathbf{C}_2^{-1})}\) into~\eqref{eq:cov_nwpm_der2} and simplifying, we derive
\begin{align}
    {\rm cov}(\hat{\boldsymbol{\theta}}_{\mathrm{tswpm}}-\boldsymbol{\theta}) =\frac{1}{\left({\rm trace}(\mathbf{C}_2^{-1})\right)^2}\left(\mathbf{H}_2^T\mathbf{C}_2^{-1}\mathbf{H}_2\right).
    \label{eq:cov_nwpm_der3}
\end{align}

The accuracy of the TS-WPM is determined by the MSE associated with \( \hat{\boldsymbol{\theta}}_{\mathrm{tswpm}} \) defined in~\eqref{eq:nwpm_final_est}. Using the result in~\eqref{eq:cov_nwpm_der3}, the MSE of TS-WPM estimate is given by
\begin{align}
    {\rm MSE}_{\mathrm{tswpm},r_1} = \frac{1}{\left({\rm trace}(\mathbf{C}_2^{-1})\right)^2}{\rm trace}\left(\mathbf{H}_2^T\mathbf{C}_2^{-1}\mathbf{H}_2\right).
    \label{eq:mse_tswpm_der1}
\end{align}

Since our desired MSE is for the position parameter 
\([x,y]\), we use the Schur complement in~\eqref{eq:efim_tdoa_scen2} and derive the MSE for \([x,y]\) as  
\begin{align}  
    {\rm MSE}_{\mathrm{tswpm}} =\frac{1}{\left({\rm trace}(\mathbf{C}_2^{-1})\right)^2}{\rm trace}\left(\mathbf{F}_{2,[x,y]}\right).  
    \label{eq:mse_tswpm_der2}  
\end{align}  

The expression in~\eqref{eq:mse_tswpm_der2} can be rewritten in terms of eigenvalues as  
\begin{align}  
    {\rm MSE}_{\mathrm{tswpm}} =\frac{1}{\left({\rm trace}(\mathbf{C}_2^{-1})\right)^2}\sum_{i=1}^K\lambda_i\left(\mathbf{F}_{2,[x,y]}\right),  
    \label{eq:mse_tswpm_der3}  
\end{align}  
where \( \lambda_i \) are the eigenvalues of \( \mathbf{F}_{2,[x,y]} \), and \( K \) denotes the number of non-zero eigenvalues, which takes a maximum value of 2 in a 2D estimation scenario.

\subsection{Proof of Theorem~\ref{thm::optimality::tswpm}}\label{app::optimality::tswpm}
Considering a two-stage estimation method of the form 
\begin{align}
    \hat{\boldsymbol{\phi}}_1& = \frac{1}{N}\sum_m h_1(\boldsymbol{\phi}_1,\mathbf{x}_m),\notag\\
    \hat{\boldsymbol{\phi}}_2& = \frac{1}{N}\sum_n h_2(\boldsymbol{\phi}_2,\hat{\boldsymbol{\phi}}_1,\mathbf{y}_n).\label{eq:two_stage_est}
\end{align}
where \(\boldsymbol{\phi}_1\) and \(\boldsymbol{\phi}_2\) are the parameters to be estimated in the first and second stages, respectively; \(h_1\) and \(h_2\) are the corresponding function maps, while \(\mathbf{x}_m\) and \(\mathbf{y}_n\) represent the respective received signals. According to ~\cite{murphy_topel,mandic_train}, for any two-stage estimation methods of this form, we can represent the error covariance of the second-stage estimate as a summation of the error covariance assuming no error propagation from the first stage to the second stage, i.e., with accurate \(\boldsymbol{\phi}_1\), and a correction term due to the erroneous estimation of \(\boldsymbol{\phi}_1\), denoted as \(\hat{\boldsymbol{\phi}}_1\). We can verify that the proposed approach outlined in Algorithm~\ref{alg:novel_ts_wpm_tdoa} aligns with~\eqref{eq:two_stage_est}, where \(\boldsymbol{\phi}_1 = r_1\), \(\boldsymbol{\phi}_2 = \boldsymbol{\theta}\), and \(\tilde{\mathbf{r}}\) is the received signal in both stages. With that understanding, we now derive the error covariance by applying MLE.

Revisiting the TDOA measurements definition in~\eqref{eq:noisy_tdoa} and rewriting it, we have
\begin{align}
\tilde{r}_b = \hat{r}_b - \hat{r}_1 = r_b - r_1 + n_b - n_1,\quad b \in \{2, 3, \cdots, B\},\label{eq:tdoa_measure_def}
\end{align}
where \(r_b = \|\boldsymbol{\theta}-\mathbf{a}_b\|\) is the true range between anchor \(b\) and the target UE and \(n_b\) is AWGN with variance \(\sigma_b^2\). Subsequently, we can express a vector of TDOA measurements as
\begin{align}
\tilde{\mathbf{r}} = \mathbf{m}(r_1,\boldsymbol{\theta}) + \mathbf{n},\label{eq:tdoa_vec_def}
\end{align}
where \(\mathbf{m}(r_1,\boldsymbol{\theta}) = \begin{bmatrix}r_2(\boldsymbol{\theta})-r_1\, \cdots\, r_B(\boldsymbol{\theta})-r_1\end{bmatrix}^T\) and the resultant noise is denoted by \(\mathbf{n}\), which has covariance \(\mathbf{C}_1\) as given in~\eqref{eq:cov_tdoa_scen1}. Then, if we fix \(\boldsymbol{\theta}\) (say at a previous estimate denoted by \(\boldsymbol{\theta}_0\)) and view \(r_1\) as the only unknown, the log-likelihood function for the first stage is formulated as
\begin{align}
L_1(r_1) = -\frac{1}{2}\Bigl(\tilde{\mathbf{r}}-\mathbf{m}(r_1,\boldsymbol{\theta}_0)\Bigr)^T \mathbf{C}_1^{-1}\Bigl(\tilde{\mathbf{r}}-\mathbf{m}(r_1,\boldsymbol{\theta}_0)\Bigr).
\end{align}

Alternatively, treating both \(\boldsymbol{\theta}\) and \(r_1\) as unknowns, we formulate the log-likelihood function for the second stage estimation as
\begin{align}
L_2(r_1,\boldsymbol{\theta}) = -\frac{1}{2}\Bigl(\tilde{\mathbf{r}}-\mathbf{m}(r_1,\boldsymbol{\theta})\Bigr)^T \mathbf{C}_1^{-1}\Bigl(\tilde{\mathbf{r}}-\mathbf{m}(r_1,\boldsymbol{\theta})\Bigr).
\end{align}

Then, we obtain the following matrices
\begin{align}
\mathbf{R}_1 &= -E\Biggl\{\frac{\partial^2 L_1}{\partial r_1^2}\Biggr\},\,\mathbf{R}_2 = -E\Biggl\{\frac{\partial^2 L_2}{\partial \boldsymbol{\theta}\,\partial\boldsymbol{\theta}^T}\Biggr\}\,, \label{eq:R1_R2}\\
\mathbf{R}_3 &= -E\Biggl\{\frac{\partial^2 L_2}{\partial \boldsymbol{\theta}\,\partial r_1}\Biggr\},\,
\mathbf{R}_4 = E\Biggl\{\frac{\partial L_1}{\partial r_1}\frac{\partial L_2}{\partial \boldsymbol{\theta}^T}\Biggr\}. \label{eq:R3_R4}
\end{align}

A standard derivation yields
\begin{align}
\mathbf{R}_1 &= \mathbf{u}^T \mathbf{C}_1^{-1}\mathbf{u},\,\mathbf{R}_2 = \mathbf{J}^T \mathbf{C}_1^{-1}\mathbf{J}\,,\notag\\
\mathbf{R}_3 &= - \mathbf{J}^T \mathbf{C}_1^{-1}\mathbf{u},\,
\mathbf{R}_4 = - \mathbf{u}^T \mathbf{C}_1^{-1}\mathbf{J}.\notag
\end{align}
where \(\mathbf{u}\) is a vector of \(B-1\) ones and \(\mathbf{J}\) is the Jacobian transformation of \(\tilde{\mathbf{r}}\) with respect to \(\boldsymbol{\theta}\). Note that \(\mathbf{R}_3=\mathbf{R}_4\). Then, according to~\cite{murphy_topel}, it can be shown that the error covariance of the second stage estimate, denoted by \(\boldsymbol{\Sigma}_{\boldsymbol{\theta}}^{\rm MLE}\), is given by
\begin{align}
\boldsymbol{\Sigma}_{\boldsymbol{\theta}}^{\rm MLE} = \mathbf{R}_2^{-1} + \boldsymbol{\Sigma}_{C}.
\label{eq:Sigma_alt}
\end{align}
where \(\boldsymbol{\Sigma}_{C}\) is a correction term introduced due to the first-stage estimate and is given by 
\begin{align}
    \boldsymbol{\Sigma}_{C} = \mathbf{R}_2^{-1}\left(\mathbf{R}_3^T \mathbf{R}_1^{-1} \mathbf{R}_3 - \mathbf{R}_4^T \mathbf{R}_1^{-1} \mathbf{R}_3 - \mathbf{R}_3^T \mathbf{R}_1^{-1} \mathbf{R}_4\right)\mathbf{R}_2^{-1}.
\end{align}

Using \(\mathbf{R}_3=\mathbf{R}_4=-\mathbf{J}^T \mathbf{C}_1^{-1}\mathbf{u}\), one can show that the expression in (\ref{eq:Sigma_alt}) simplifies to
\begin{align}
\boldsymbol{\Sigma}_{\boldsymbol{\theta}}^{\rm MLE}=\left(\mathbf{J}^T \mathbf{C}_1^{-1}\mathbf{J}-\frac{\mathbf{J}^T \mathbf{C}_1^{-1}\mathbf{u}\,\mathbf{u}^T \mathbf{C}_1^{-1}\mathbf{J}}{\mathbf{u}^T \mathbf{C}_1^{-1}\mathbf{u}}\right)^{-1}.
\end{align}

As derived earlier, the TS-WPM estimator yields an error covariance (as specified in~\eqref{eq:cov_nwpm_der3})
\begin{align}
{\rm cov}(\hat{\boldsymbol{\theta}}_{\mathrm{tswpm}}-\boldsymbol{\theta}) =\frac{1}{\left({\rm trace}(\mathbf{C}_2^{-1})\right)^2}\left(\mathbf{H}_2^T\mathbf{C}_2^{-1}\mathbf{H}_2\right).
\label{eq:cov_ts_wpm}
\end{align}

TS-WPM approaches MLE, i.e., \({\rm cov}(\hat{\boldsymbol{\theta}}_{\mathrm{tswpm}}-\boldsymbol{\theta})\) becomes equal to \(\boldsymbol{\Sigma}_{\boldsymbol{\theta}}^{\rm MLE}\) when the cross-information between \(r_1\) and \(\boldsymbol{\theta}\) vanishes, i.e., \(\mathbf{u}^T \mathbf{C}_1^{-1}\mathbf{J} = \mathbf{0}\). This condition is met when the measurement noises are uncorrelated. This situation occurs when the covariance matrix \(\mathbf{C}_1\) reduces to \(\mathbf{C}_2\), which occurs when the reference anchor does not introduce additional error dependencies--in practice when its SNR approaches infinity (i.e., \(\sigma_1^2 \rightarrow 0\)). In other words, in a high SNR regime, the proposed novel TS-WPM is {\em asymptotically optimal}.

\subsection{Proof of Lemma~\ref{lemma::cov::mse::wnls}}\label{app::cov::mse::wnls}
Let \( \hat{\boldsymbol{\theta}}^k \) represent the iterative update of \( \boldsymbol{\theta} \) at iteration \( k \). As described in~\cite{buehrer2019handbook,kay1993fundamentals}, the WNLS update is expressed as
\begin{align}
\hat{\boldsymbol{\theta}}^k = \hat{\boldsymbol{\theta}}^{k-1} + \boldsymbol{\beta}_{k-1}\big(\mathbf{H}_1(\hat{\boldsymbol{\theta}}^{k-1})\big)^T\mathbf{W}_1\big(\tilde{\mathbf{r}} - \tilde{\mathbf{r}}_e(\hat{\boldsymbol{\theta}}^{k-1})\big),
\label{eq:weighted_nls_def}
\end{align}
where \( \boldsymbol{\beta}_{k-1}=\big((\mathbf{H}_1(\hat{\boldsymbol{\theta}}^{k-1}))^T\mathbf{W}_1\mathbf{H}_1(\hat{\boldsymbol{\theta}}^{k-1})\big)^{-1} \), and \( \mathbf{W}_1 \) is defined as \( \mathbf{W}_1=\mathbf{C}_1^{-1}\), with \( \mathbf{C}_1 \) given in~\eqref{eq:cov_tdoa_scen1}. The term \( \tilde{\mathbf{r}}_e(\hat{\boldsymbol{\theta}}^{k-1}) \) represents a vector of TDOA estimates evaluated at \( \hat{\boldsymbol{\theta}}^{k-1} \) as given in~\eqref{eq:tdoa_est_method1}.

Assuming convergence, the final WNLS estimate can be approximated as
\begin{align}
\hat{\boldsymbol{\theta}}_{\mathrm{wnls}} = \boldsymbol{\theta} + \boldsymbol{\beta}(\mathbf{H}_1(\boldsymbol{\theta}))^T\mathbf{W}_1\big(\tilde{\mathbf{r}} - \tilde{\mathbf{r}}_e(\boldsymbol{\theta})\big),
\label{eq:weighted_nls_final_est}
\end{align}
where \( \boldsymbol{\theta} \) represents the true position parameter, \( \mathbf{H}_1(\boldsymbol{\theta}) \) denotes the Jacobian matrix of \( \tilde{\mathbf{r}} \) with respect to \( \boldsymbol{\theta} \), equal to \( \mathbf{H}_1 \) as specified in~\eqref{eq:jacobian_tdoa}, and \( \boldsymbol{\beta}=\left((\mathbf{H}_1(\boldsymbol{\theta}))^T\mathbf{W}_1\mathbf{H}_1(\boldsymbol{\theta})\right)^{-1}\). In the following derivation, we will use \(\mathbf{H}_1\) instead of \(\mathbf{H}_1(\boldsymbol{\theta}) \).

The MSE associated with the estimate \( \hat{\boldsymbol{\theta}}_{\mathrm{wnls}} \) quantifies the positioning error of WNLS. To derive the MSE, we first determine the covariance of \( (\hat{\boldsymbol{\theta}}_{\mathrm{wnls}}-\boldsymbol{\theta}) \) as follows
\begin{align}
    {\rm cov}(\hat{\boldsymbol{\theta}}_{\mathrm{wnls}}-\boldsymbol{\theta}) = \boldsymbol{\beta}\mathbf{H}_1^T\mathbf{W}_1{\rm cov}(\tilde{\mathbf{r}} - \tilde{\mathbf{r}}_e(\boldsymbol{\theta}))\mathbf{W}_1\mathbf{H}_1\boldsymbol{\beta}^T,
    \label{eq:cov_wnls_der1}
\end{align}
where \( {\rm cov}(\tilde{\mathbf{r}} - \tilde{\mathbf{r}}_e(\boldsymbol{\theta})) \) is essentially \( \mathbf{C}_1 \) as defined in~\eqref{eq:cov_tdoa_scen1}. Substituting \( \mathbf{C}_1 \) into~\eqref{eq:cov_wnls_der1} yields
\begin{align}
    {\rm cov}(\hat{\boldsymbol{\theta}}_{\mathrm{wnls}}-\boldsymbol{\theta}) =\boldsymbol{\beta}\mathbf{H}_1^T\mathbf{W}_1\mathbf{C}_1\mathbf{W}_1\mathbf{H}_1\boldsymbol{\beta}^T.
    \label{eq:cov_wnls_der2}
\end{align}

Further, substituting \( \boldsymbol{\beta} \) and \( \mathbf{W}_1 \) into~\eqref{eq:cov_wnls_der2} and simplifying leads to
\begin{align}
    {\rm cov}(\hat{\boldsymbol{\theta}}_{\mathrm{wnls}}-\boldsymbol{\theta}) =\left(\mathbf{H}_1^T\mathbf{C}_1^{-1}\mathbf{H}_1\right)^{-1}.
    \label{eq:cov_wnls_der3}
\end{align}

Consequently, the MSE associated with the WNLS estimate is
\begin{align}
    {\rm MSE}_{\mathrm{wnls}} = {\rm trace}\left(\left(\mathbf{H}_1^T\mathbf{C}_1^{-1}\mathbf{H}_1\right)^{-1}\right).
    \label{eq:mse_wnls_der1}
\end{align}

The expression in~\eqref{eq:mse_wnls_der1} can be expressed in terms of eigenvalues as
\begin{align}
    {\rm MSE}_{\mathrm{wnls}} = \sum_{i=1}^K \frac{1}{\lambda_i\left(\mathbf{H}_1^T\mathbf{C}_1^{-1}\mathbf{H}_1\right)},
    \label{eq:mse_wnls_der2}
\end{align}
where \( \lambda_i \) are the eigenvalues of \( \mathbf{H}_1^T\mathbf{C}_1^{-1}\mathbf{H}_1 \), and \( K \) indicates the number of non-zero eigenvalues. For 2D estimation, the maximum value of \( K \) is 2.

\subsection{Proof of Proposition~\ref{prop::robustness::tswpm}}\label{app::robustness::tswpm}
Considering the MSE of WNLS given in~\eqref{eq:mse_wnls_der2}, in high GDOP scenarios, for 2D estimation, one eigenvalue of \( \mathbf{H}_1^T\mathbf{C}_1^{-1}\mathbf{H}_1 \) is quite significant, while the other remains relatively insignificant. Consequently, the MSE in ~\eqref{eq:mse_wnls_der2} can be approximated as
\begin{align}
    {\rm MSE}_{\mathrm{wnls}} \approx \frac{1}{\lambda_{\min}\left(\mathbf{H}_1^T\mathbf{C}_1^{-1}\mathbf{H}_1\right)}.
    \label{eq:mse_wnls_final}
\end{align}
 




Similarly, we approximate the MSE of TS-WPM given in~\eqref{eq:mse_tswpm_der3} in high GDOP scenarios. To facilitate this, we first simplify the second term in~\eqref{eq:efim_tdoa_scen2}. We denote the expression \(\mathbf{C}_2^{-1}\mathbf{v}\left(\mathbf{v}^T\mathbf{C}_2^{-1}\mathbf{v}\right)^{-1}\mathbf{v}^T\mathbf{C}_2^{-1}\) in~\eqref{eq:efim_tdoa_scen2} as \(\mathbf{C_v}\) and evaluate it to obtain
\begin{align}
    \mathbf{C_v} = \left(\sum_{b=1}^B\frac{1}{\sigma_b^2}\right)\begin{bmatrix}
        \left(\frac{1}{\sigma_1^2}\right)^2 & \frac{1}{\sigma_1^2\sigma_2^2} & \cdots & \frac{1}{\sigma_1^2\sigma_B^2}\\
        \vdots & \vdots & \ddots & \vdots\\
        \frac{1}{\sigma_B^2\sigma_1^2} & \frac{1}{\sigma_B^2\sigma_2^2} & \cdots & \left(\frac{1}{\sigma_B^2}\right)^2
    \end{bmatrix}.
    \label{eq:term2_efim_tdoa_scen2}
\end{align}

In high GDOP scenarios, when the anchors are colocated, we observe similar SNR across all the anchors and the target UE links, resulting in identical \(\sigma_b^2\) values for all the anchors. This behavior causes \(\mathbf{C_v}\) to have identical entries. Consequently, we can verify that \({\rm trace}\left(\mathbf{H}_t^T\mathbf{C_v}\mathbf{H}_t\right)\) approaches 0. Using this result, we can show that in high GDOP scenarios, 
\begin{align}
    {\rm trace}\left(\mathbf{F}_{2,[x,y]}\right)\approx{\rm trace}\left(\mathbf{H}_t^T\mathbf{C}_2^{-1}\mathbf{H}_t\right).
\end{align}

By using this result along with \({\rm trace}\left(\mathbf{H}_t^T\mathbf{C}_2^{-1}\mathbf{H}_t\right)={\rm trace}\left(\mathbf{C}_2^{-1}\right)\), the MSE in~\eqref{eq:mse_tswpm_der2} can be approximated for high GDOP scenarios as
\begin{align}
    {\rm MSE}_{\mathrm{tswpm}} \approx \frac{1}{\sum_{b=1}^B\frac{1}{\sigma_b^2}}.
    \label{eq:mse_tswpm_final}
\end{align}

This result is significant, as it shows that even under high GDOP conditions, the MSE of the proposed two-stage WPM remains directly proportional to the variance of pseudorange measurements, which is not commonly observed in non-ideal settings~\cite{buehrer2019handbook}. By comparing~\eqref{eq:mse_wnls_final} and~\eqref{eq:mse_tswpm_final}, we observe that under high GDOP, \( \lambda_{\min}\left(\mathbf{H}_1^T\mathbf{C}_1^{-1}\mathbf{H}_1\right) \) becomes insignificant, resulting in high MSE for the WNLS case. In contrast, the MSE with the novel TS-WPM in~\eqref{eq:mse_tswpm_final} trends similarly to low GDOP or favorable conditions, maintaining robust performance and reinforcing its effectiveness in non-ideal situations.
\balance
\bibliographystyle{IEEEtran}
\bibliography{hokie}
\vskip -1\baselineskip plus -1fil
\begin{IEEEbiography}[{\includegraphics[width=1in,height=1.25in,clip,keepaspectratio]{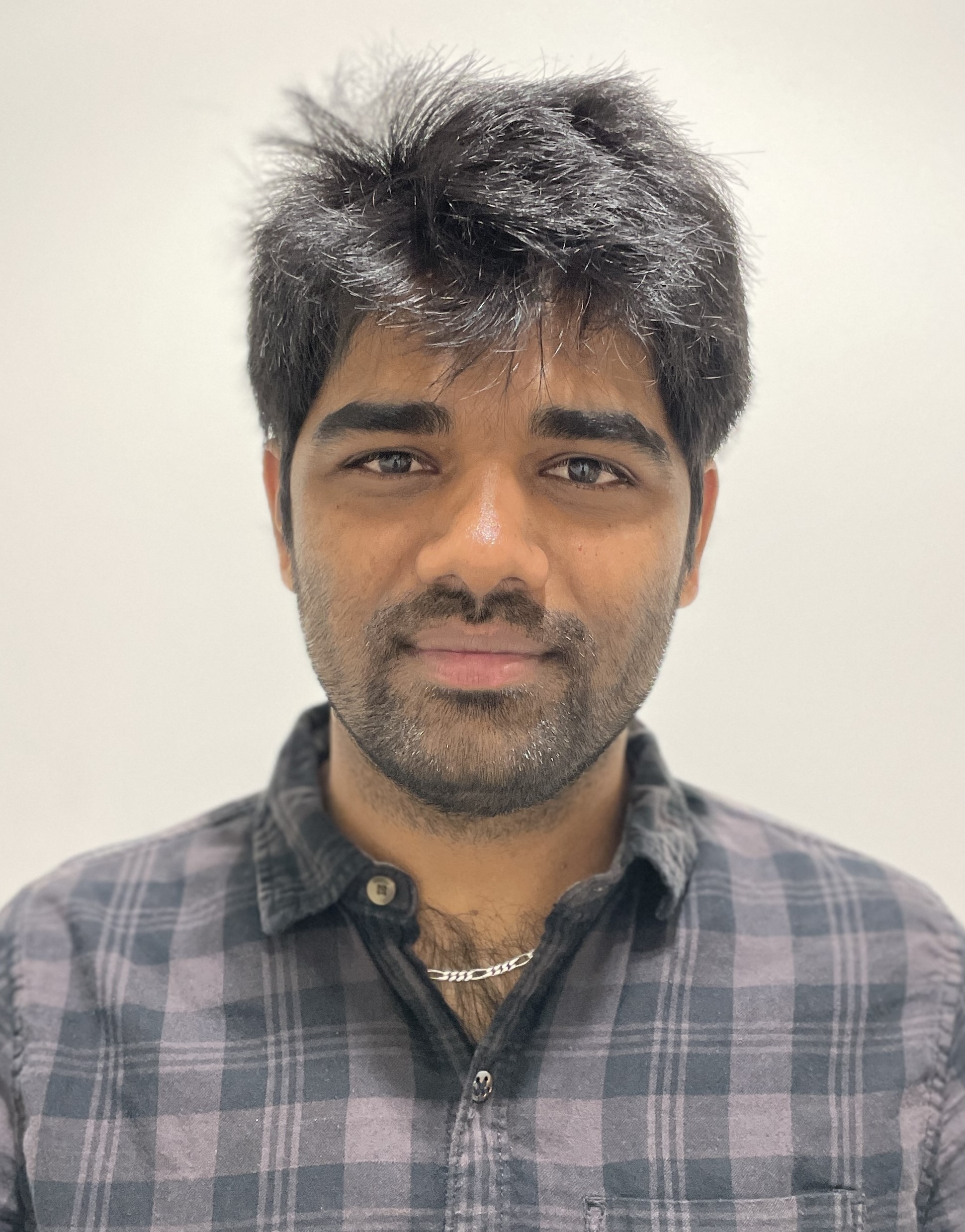}}]{Harish K. Dureppagari}
(Graduate Student Member, IEEE) received his B.Tech. degree in Electronics and Communication Engineering from Rajiv Gandhi University of Knowledge Technologies (RGUKT), India, in 2014, and his M.Tech. degree in Electrical Engineering from the Indian Institute of Technology (IIT) Hyderabad, India, in 2016. Since Fall 2022, he has been pursuing the Ph.D. degree with the Bradley Department of Electrical and Computer Engineering at Virginia Tech, Blacksburg, VA, USA. He has held summer internship positions at Nokia, IL, USA (Summer 2023), and Qualcomm, San Diego, CA, USA (Summers 2024 and 2025). Prior to his Ph.D. studies, he worked as a Lead Engineer at WiSig Networks Pvt. Ltd., India. His research interests include wireless communications, localization, and massive MIMO.
\end{IEEEbiography}
\vskip 0\baselineskip plus -10fil
\begin{IEEEbiography} [{\includegraphics[width=1in,height=1.25in,clip]{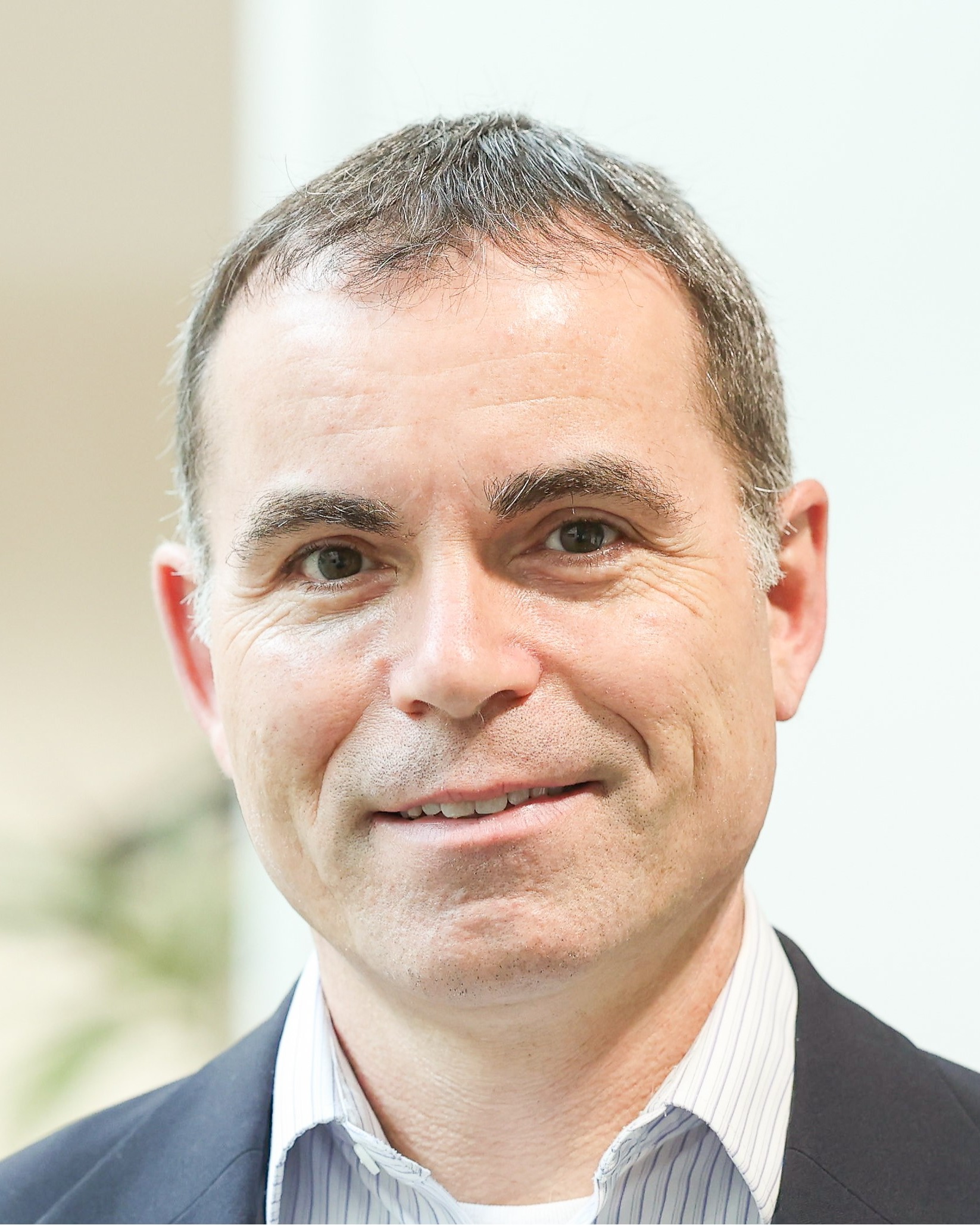}}]{R. Michael Buehrer}
(Fellow, IEEE) joined Virginia Tech from Bell Labs as an Assistant Professor with the Bradley Department of Electrical and Computer Engineering in 2001.  He is currently a Professor and the Fred C. Lee Endowed Chair of Electrical and Computer Engineering and is the former Director of Wireless @ Virginia Tech, a comprehensive research group focusing on wireless communications, radar and localization.  During 2009 Dr. Buehrer was a visiting researcher at the Laboratory for Telecommunication Sciences (LTS) a federal research lab which focuses on telecommunication challenges for national defense.  While at LTS, his research focus was in the area of cognitive radio with a particular emphasis on statistical learning techniques.

Dr. Buehrer was named an IEEE Fellow in 2016 ``for contributions to wideband signal processing in communications and geolocation."  In 2023, he received the prestigious MILCOM Lifetime Award for Technical Achievement. This award recognizes individuals who have made important technical contributions to military communications over the course of their careers. His current research interests include machine learning for wireless communications and radar, geolocation, position location networks, cognitive radio, cognitive radar, electronic warfare, dynamic spectrum sharing, communication theory, Multiple Input Multiple Output (MIMO) communications, spread spectrum, interference avoidance, and propagation modeling.  His work has been funded by the National Science Foundation, the Defense Advanced Research Projects Agency, the Office of Naval Research, the Army Research Office, the Air Force Research Lab and several industrial sponsors.

Dr. Buehrer has authored or co-authored over 100 journal and approximately 300 conference papers and holds 18 patents in the area of wireless communications.  In 2023, he received the prestigious MILCOM Lifetime Award for Technical Achievement, an award that recognizes individuals who have made important technical contributions to military communications over the course of their careers.  In 2023 and 2021 he was the co-recipient of the Vanu Bose Award for the best paper at IEEE MILCOM.  In 2023 and 2010 he was co-recipient of the Fred W. Ellersick MILCOM Award for the best paper in the unclassified technical program.  He was formerly an Area Editor IEEE Wireless Communications.  He was also formerly an associate editor for IEEE Transactions on Communications, IEEE Transactions on Vehicular Technologies, IEEE Transactions on Wireless Communications, IEEE Transactions on Signal Processing, IEEE Wireless Communications Letters, and IEEE Transactions on Education.  He has also served as a guest editor for special issues of The Proceedings of the IEEE, and IEEE Transactions on Special Topics in Signal Processing.  In 2003 he was named Outstanding New Assistant Professor by the Virginia Tech College of Engineering and in 2014 he received the Dean's Award for Excellence in Teaching.
\end{IEEEbiography}
\vskip 0\baselineskip plus -1fil
\begin{IEEEbiography} [{\includegraphics[width=1in,height=1.25in,clip,keepaspectratio]{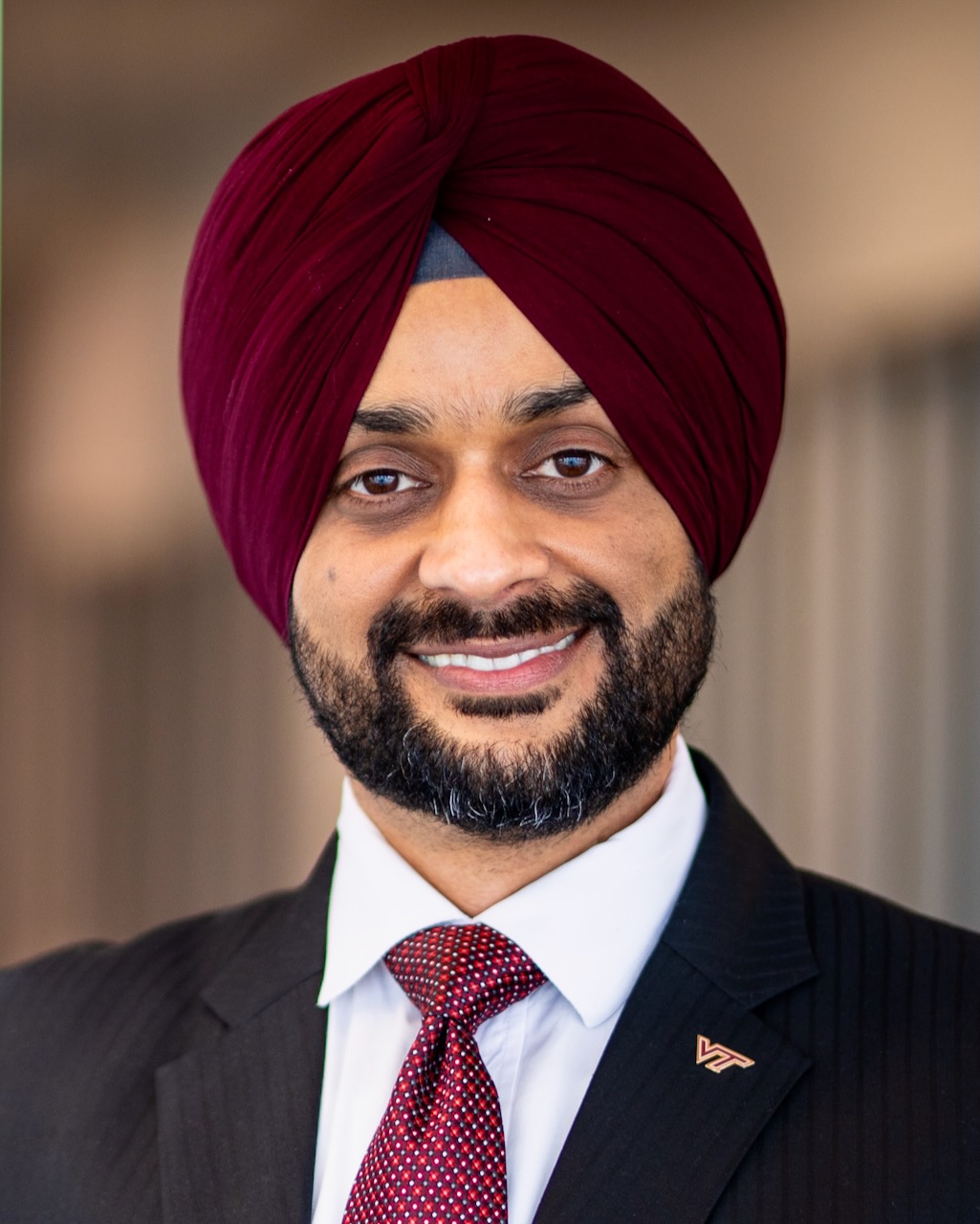}}]{Harpreet S. Dhillon}
(Fellow, IEEE) received his B.Tech. from IIT Guwahati in 2008, his M.S. from Virginia Tech in 2010, and his Ph.D. from the University of Texas at Austin in 2013.  

Following a year as a Viterbi Postdoctoral Fellow at the University of Southern California, he joined Virginia Tech in 2014, where he is currently the W. Martin Johnson Professor of Engineering and Associate Dean for Research and Innovation. In 2024, he also served as the Interim Department Head of Electrical and Computer Engineering (ECE). His research interests span communication theory, wireless networks, geolocation, and stochastic geometry. He is a Fellow of AAIA and AIIA, a Clarivate Analytics Highly Cited Researcher, and a recipient of six best paper awards, including the IEEE Leonard G. Abraham Prize (2014), the IEEE Katherine Johnson Young Author Best Paper Award (2015), and the IEEE Heinrich Hertz Award (2016). He has also received early-career technical achievement awards from three IEEE ComSoc Technical Committees. At Virginia Tech, his honors include the Outstanding New Assistant Professor Award (2017), the Dean's Award for Excellence in Research (2020), and multiple named faculty fellowships, including the Steven O. Lane Junior Faculty Fellowship (2018), the College of Engineering Faculty Fellowship (2018), and the Elizabeth and James E. Turner Jr. '56 Faculty Fellowship (2019). His other recognitions include the Agilent Engineering and Technology Award (2008), the UT Austin MCD Fellowship, the UT Austin WNCG Leadership Award (2013), and the inaugural IIT Guwahati Young Alumni Achiever Award (2020). He has served as TPC Co-chair for IEEE WCNC 2022 and IEEE PIMRC 2024, as symposium TPC Co-chair for several IEEE conferences, and on the editorial boards of multiple IEEE journals. He is currently a member of the Executive Editorial Committee of the {\sc IEEE Transactions on Wireless Communications}.
\end{IEEEbiography}
\end{document}

%% file: notation.tex
\def\nb0{{\mathbf{0}}}
\def\nb1{{\mathbf{1}}}

\newtheorem{lemma}{Lemma}

\newtheorem{theorem}{Theorem}
\newtheorem{corollary}{Corollary}[theorem] 
\newtheorem{prop}{Proposition}










%% file: revised_main_draft.bbl
\begin{thebibliography}{10}
\providecommand{\url}[1]{#1}
\csname url@samestyle\endcsname
\providecommand{\newblock}{\relax}
\providecommand{\bibinfo}[2]{#2}
\providecommand{\BIBentrySTDinterwordspacing}{\spaceskip=0pt\relax}
\providecommand{\BIBentryALTinterwordstretchfactor}{4}
\providecommand{\BIBentryALTinterwordspacing}{\spaceskip=\fontdimen2\font plus
\BIBentryALTinterwordstretchfactor\fontdimen3\font minus \fontdimen4\font\relax}
\providecommand{\BIBforeignlanguage}[2]{{%
\expandafter\ifx\csname l@#1\endcsname\relax
\typeout{** WARNING: IEEEtran.bst: No hyphenation pattern has been}%
\typeout{** loaded for the language `#1'. Using the pattern for}%
\typeout{** the default language instead.}%
\else
\language=\csname l@#1\endcsname
\fi
#2}}
\providecommand{\BIBdecl}{\relax}
\BIBdecl

\bibitem{dureppagari2025icc}
H.~K. Dureppagari, R.~M. Buehrer, and H.~S. Dhillon, ``{A Novel Positioning Framework: Two-Stage Weighted Projection with NLOS Bias Modeling},'' in \emph{Proc. IEEE ICC Workshops}, 2025, pp. 1476--1481.

\bibitem{kay1993fundamentals}
S.~M. Kay, ``{Fundamentals of Statistical Signal Processing: Estimation Theory},'' 1993.

\bibitem{buehrer2019handbook}
R.~Zekavat and R.~M. Buehrer, ``{Handbook of Position Location: Theory, Practice, and Advances},'' 2019.

\bibitem{9376594}
S.~Zhao, X.-P. Zhang, X.~Cui, and M.~Lu, ``{Semidefinite Programming Two-Way TOA Localization for User Devices With Motion and Clock Drift},'' \emph{IEEE Signal Process. Lett.}, vol.~28, pp. 578--582, 2021.

\bibitem{6268731}
R.~M. Vaghefi and R.~M. Buehrer, ``{Cooperative sensor localization with NLOS mitigation using semidefinite programming},'' in \emph{IEEE WPNC}, 2012, pp. 13--18.

\bibitem{8241388}
Z.~Su, G.~Shao, and H.~Liu, ``{Semidefinite Programming for NLOS Error Mitigation in TDOA Localization},'' \emph{IEEE Commun. Lett.}, vol.~22, no.~7, pp. 1430--1433, 2018.

\bibitem{10.1145/1138127.1138129}
J.~A. Costa, N.~Patwari, and A.~O. Hero, ``{Distributed weighted-multidimensional scaling for node localization in sensor networks},'' \emph{ACM Trans. Sen. Netw.}, vol.~2, no.~1, pp. 39--64, Feb. 2006.

\bibitem{5342507}
H.-W. Wei, R.~Peng, Q.~Wan, Z.-X. Chen, and S.-F. Ye, ``{Multidimensional Scaling Analysis for Passive Moving Target Localization With TDOA and FDOA Measurements},'' \emph{IEEE Trans. Signal Process.}, vol.~58, no.~3, pp. 1677--1688, 2010.

\bibitem{Bhatti2018MachineLB}
G.~M. Bhatti, ``{Machine Learning Based Localization in Large-Scale Wireless Sensor Networks},'' \emph{Sensors}, vol.~18, 2018.

\bibitem{tian2023highprecisionmachinelearningbasedindoor}
G.~Tian, I.~Yaman, M.~Sandra, X.~Cai, L.~Liu, and F.~Tufvesson, ``{High-Precision Machine-Learning Based Indoor Localization with Massive MIMO System},'' \emph{arXiv:2303.03743}, 2023.

\bibitem{5683693_toa_coop_mod_ippm}
T.~Jia and R.~M. Buehrer, ``{Collaborative Position Location for Wireless Networks Using Iterative Parallel Projection Method},'' in \emph{IEEE GLOBECOM}, 2010, pp. 1--6.

\bibitem{5677549_tao_coop_ippm}
------, ``{A Set-Theoretic Approach to Collaborative Position Location for Wireless Networks},'' \emph{IEEE Trans. Mobile Comput.}, vol.~10, no.~9, pp. 1264--1275, 2011.

\bibitem{10632801_harish_dyspan}
H.~K. Dureppagari, C.~Saha, H.~S. Dhillon, and R.~M. Buehrer, ``{UAV-Based 5G Localization for Emergency Response Using Signals of Opportunity},'' in \emph{Proc. IEEE DySPAN Workshops}, 2024, pp. 63--68.

\bibitem{10329418}
H.~Kwon, C.~Hegde, Y.~Kiarashi, V.~S.~K. Madala, R.~Singh, A.~Nakum, R.~Tweedy, L.~M. Tonetto, C.~M. Zimring, M.~Doiron, A.~D. Rodriguez, A.~I. Levey, and G.~D. Clifford, ``{A Feasibility Study on Indoor Localization and Multiperson Tracking Using Sparsely Distributed Camera Network With Edge Computing},'' \emph{IEEE J-ISPIN}, vol.~1, pp. 187--198, 2023.

\bibitem{10.1007/s00521-021-06815-9}
X.~Zhang, F.~He, Q.~Chen, X.~Jiang, J.~Bao, T.~Ren, and X.~Du, ``{A differentially private indoor localization scheme with fusion of WiFi and bluetooth fingerprints in edge computing},'' \emph{Neural Comput. Appl.}, vol.~34, no.~6, pp. 4111--4132, Mar. 2022.

\bibitem{dureppagari_ntn_10355106}
H.~K. Dureppagari, C.~Saha, H.~S. Dhillon, and R.~M. Buehrer, ``{NTN-Based 6G Localization: Vision, Role of LEOs, and Open Problems},'' \emph{IEEE Wirel. Commun.}, vol.~30, no.~6, pp. 44--51, 2023.

\bibitem{dureppagari2024leo}
H.~K. Dureppagari, C.~Saha, H.~Krishnamurthy, X.~F. Wang, A.~Rico-Alvari{\~n}o, R.~M. Buehrer, and H.~S. Dhillon, ``{LEO-based Positioning: Foundations, Signal Design, and Receiver Enhancements for 6G NTN},'' \emph{IEEE Commun. Mag.}, to appear.

\bibitem{10634076}
F.~Munier, Z.~Xiong, R.~Shreevastav, X.~Jiang, Y.~Lyazidi, D.~Shrestha, C.~Zhang, G.~Lindmark, J.~Nygren, S.~Moloudi, S.~Dwivedi, and F.~Gunnarsson, ``{Positioning of RedCap Devices in 5G Networks},'' \emph{IEEE Commun. Mag.}, vol.~62, no.~8, pp. 110--116, 2024.

\bibitem{ZHUO2024}
Y.~Zhuo and Z.~Wang, ``{Low-complexity multi-target localization via multi-BS sensing},'' \emph{Digit. Commun. Netw.}, 2024.

\bibitem{5464262}
J.~Yan, C.~C. J.~M. Tiberius, P.~J.~G. Teunissen, G.~Bellusci, and G.~J.~M. Janssen, ``{A Framework for Low Complexity Least-Squares Localization With High Accuracy},'' \emph{IEEE Trans. Signal Process.}, vol.~58, no.~9, pp. 4836--4847, 2010.

\bibitem{8947090_liu_weight_ippm}
T.~Liu, S.~Tian, G.~Li, L.~Lu, Y.~Tian, and W.~Dai, ``{Cooperative Positioning via Weighted Parallel Projection for Wireless Sensor Networks},'' in \emph{IEEE ICCT}, 2019, pp. 440--445.

\bibitem{8979252_liu_covar_ippm}
T.~Liu, G.~Li, L.~Lu, S.~Li, and S.~Tian, ``{Robust Hybrid Cooperative Positioning Via a Modified Distributed Projection-Based Method},'' \emph{IEEE Trans. Wirel. Commun.}, vol.~19, no.~5, pp. 3003--3018, 2020.

\bibitem{3gpp::38211}
``{Technical Specification Group Radio Access Network; NR; Physical channels and modulation (Release 18) },'' \emph{3GPP TS 38.211, version 18.0.0}, Sep. 2023.

\bibitem{3gpp::36814}
``{Technical Specification Group Radio Access Network; Evolved Universal Terrestrial Radio Access (E-UTRA); Further advancements for E-UTRA physical layer aspects (Release 9) },'' \emph{3GPP TR 36.814, version 9.2.0}.

\bibitem{3gpp::38901}
``{Technical Specification Group Radio Access Network; Study on channel model for frequencies from 0.5 to 100 GHz (Release 18) },'' \emph{3GPP TR 38.901, version 18.0.0}, Mar. 2024.

\bibitem{539767}
C.~Berrou and A.~Glavieux, ``{Near optimum error correcting coding and decoding: turbo-codes},'' \emph{IEEE Trans. Commun.}, vol.~44, no.~10, pp. 1261--1271, 1996.

\bibitem{3gpp::38355}
``{Technical Specification Group Radio Access Network; NR; Sidelink Positioning Protocol (SLPP); Protocol specification (Release 18)},'' \emph{3GPP TS 38355, version 18.4.0}, Dec. 2024.

\bibitem{9810941}
D.~Coppens, A.~Shahid, S.~Lemey, B.~Van~Herbruggen, C.~Marshall, and E.~De~Poorter, ``{An Overview of UWB Standards and Organizations (IEEE 802.15.4, FiRa, Apple): Interoperability Aspects and Future Research Directions},'' \emph{IEEE Access}, vol.~10, pp. 70\,219--70\,241, 2022.

\bibitem{ibrahim2018verification}
M.~Ibrahim, H.~Liu, M.~Jawahar, V.~Nguyen, M.~Gruteser, R.~Howard, B.~Yu, and F.~Bai, ``{Verification: Accuracy evaluation of WiFi fine time measurements on an open platform},'' in \emph{ACM MobiCom}, 2018, pp. 417--427.

\bibitem{murphy_topel}
K.~M. Murphy and R.~H. Topel, ``{Estimation and Inference in Two-Step Econometric Models},'' \emph{J. Bus. Econ. Stat.}, vol.~3, no.~4, pp. 370--379, 1985.

\bibitem{MoeYuanP12010}
Y.~Shen and M.~Win, ``Fundamental limits of wideband localization - part {I}: A general framework,'' \emph{IEEE Trans. Inf. Theory}, vol.~56, no.~10, pp. 4956 -- 4980, Oct. 2010.

\bibitem{4753258}
T.~Jia and R.~M. Buehrer, ``{A new Cramer-Rao lower bound for TOA-based localization},'' in \emph{IEEE MILCOM}, 2008, pp. 1--5.

\bibitem{QiKo2006}
Y.~Qi, H.~Kobayashi, and H.~Suda, ``{Analysis of wireless geolocation in a non-line-of-sight environment},'' \emph{IEEE Trans. Wirel. Commun.}, vol.~5, no.~3, pp. 672 -- 681, Mar. 2006.

\bibitem{6427618}
S.~Hara, D.~Anzai, T.~Yabu, K.~Lee, T.~Derham, and R.~Zemek, ``{A Perturbation Analysis on the Performance of TOA and TDOA Localization in Mixed LOS/NLOS Environments},'' \emph{IEEE Trans. Commun.}, vol.~61, no.~2, pp. 679--689, 2013.

\bibitem{dureppagari_uav1_10139944}
H.~K. Dureppagari, D.-R. Emenonye, H.~S. Dhillon, and R.~M. Buehrer, ``{UAV-Aided Indoor Localization of Emergency Response Personnel},'' in \emph{IEEE/ION PLANS}, 2023, pp. 1241--1249.

\bibitem{10.1007/s11277-007-9375-z}
S.~Gezici, ``{A Survey on Wireless Position Estimation},'' \emph{Wirel. Pers. Commun.}, vol.~44, no.~3, pp. 263--282, Feb. 2008.

\bibitem{10.5555/555289}
J.~J. Caffery, \emph{{Wireless Location in CDMA Cellular Radio Systems}}.\hskip 1em plus 0.5em minus 0.4em\relax USA: {Kluwer Academic Publishers}, 1999.

\bibitem{Edkgps2005}
E.~D. Kaplan and C.~J. Hegarty, \emph{{Understanding GPS Principles and Applications, Second Edition}}.\hskip 1em plus 0.5em minus 0.4em\relax Artech, 2005.

\bibitem{Langley1999DOP}
R.~B. Langley, ``{Dilution of Precision},'' \emph{GPS World}, vol.~10, no.~5, pp. 52--59, May 1999.

\bibitem{mandic_train}
P.~Karaca-Mandic and K.~Train, ``{Standard error correction in two-stage estimation with nested samples},'' \emph{Econom. J.}, vol.~6, no.~2, pp. 401--407, 11 2003.

\end{thebibliography}
